%% file: BEST-PPNP.tex
\documentclass[review]{elsarticle}

\usepackage{lineno}
\usepackage{tablefootnote}
\modulolinenumbers[5]

\usepackage[colorlinks,citecolor=blue,linktoc=all,linkcolor=cyan]{hyperref}
\usepackage{graphicx}

\usepackage[T1]{fontenc}
\usepackage{dsfont}               % use mathds instead of mathbb for outline fonts
\usepackage{mathrsfs}             % provides mathscr without overwriting mathcal
\usepackage{slashed}              % For Dirac slash notation.
\usepackage{amsmath}
\usepackage{amssymb}
\usepackage{amsbsy}
\usepackage{amsfonts}
\usepackage{multirow}
\usepackage{color}

\numberwithin{equation}{section}
\numberwithin{table}{section}
\numberwithin{figure}{section}

\journal{Progress in Particle and Nuclear Physics}

\usepackage{titlesec}
\usepackage{sectsty}
\titleformat{\section}{\normalfont\Large\bfseries}{\thesection}{1em}{}
\titleformat{\subsection}{\normalfont\large\bfseries}{\thesubsection}{1em}{}
\titleformat{\subsubsection}{\normalfont\normalsize\bfseries}{\thesubsubsection}{1em}{}
\input{BESTAbbreviations.tex}

\begin{document}
	
	\begin{frontmatter}
		
		\title{The Gallium Anomaly}
				
		%authors, affiliations, corresponding author mention 
		\author[mymainaddress]{Steven R. Elliott\corref{mycorrespondingauthor}}
		\cortext[mycorrespondingauthor]{Steven R. Elliott}
		\ead{elliotts@lanl.gov}

		\author[mysecondaryaddress]{Vladimir Gavrin}
		\author[mythirdaddress,myfourthaddress]{Wick Haxton}

		\address[mymainaddress]{Los Alamos National Laboratory, Los Alamos, NM 87545, USA}
		\address[mysecondaryaddress]{Institute for Nuclear Research of the Russian Academy of Sciences, Moscow 117312, Russia}
		\address[mythirdaddress]{Department of Physics, University of California, Berkeley, CA 94720, USA}
		\address[myfourthaddress]{Lawrence Berkeley National Laboratory, Berkeley, CA 94720, USA}
		
		\begin{abstract}
	        In order to test the end-to-end operations of gallium solar neutrino experiments, intense electron-capture sources
	        were fabricated to measure the responses of the radiochemical SAGE and GALLEX/GNO detectors to known fluxes of low-energy neutrinos.   Such tests were viewed 
	        at the time as a cross-check, given the many tests of \nuc{71}Ge recovery and counting that had been routinely performed, with excellent results.
	        However, the four  \nuc{51}Cr and \nuc{37}Ar source experiments yielded rates below expectations, a result commonly known as the Ga anomaly.
	        As the intensity of the electron-capture sources can be measured to high precision, the neutrino lines they produce are fixed by known atomic
	        and nuclear rates, and the neutrino absorption cross section on \nuc{71}Ga is tightly constrained by the lifetime of \nuc{71}Ge, no simple
	        explanation for the anomaly has been found.  To check these calibration experiments, a dedicated experiment BEST was performed, utilizing
	        a neutrino source of unprecedented intensity and a detector optimized to increase statistics while providing some information on
	        counting rate as a function of distance from the source.  The results BEST obtained are consistent with the earlier solar neutrino
	        calibration experiments, and when combined with those measurements, yield a Ga anomaly with a significance of approximately 4$\sigma$,
	        under conservative assumptions.  But BEST found no evidence of distance dependence and thus no explicit
	        indication of new physics.  In this review we describe the extensive campaigns carried out by SAGE, GALLEX/GNO, and BEST to 
	        demonstrate the reliability and precision of their experimental procedures, including \nuc{71}Ge recovery, counting,
	        and analysis.  We also describe efforts to define uncertainties in the neutrino capture cross section, which now include estimates
	        of effects at the $\lesssim 0.5$\% level such as radiative corrections and weak magnetism.  With the results from BEST, an anomaly remains
	        even if one retains only the transition to the \nuc{71}Ge ground state, whose strength is fixed by the known lifetime of \nuc{71}Ge.  We then consider
	        the new-physics solution most commonly suggested to resolve the Ga anomaly, oscillations into a sterile fourth neutrino, \oscil.  We find such a
	        solution generates substantial tension with several null experiments, owing to the large mixing angle required.  While this does not 
	        exclude such solutions -- the sterile sector might include multiple neutrinos as well as new interactions -- it shows the need for more 
	        experimental constraints, if we are to make progress in resolving the Ga and other low-energy neutrino anomalies.	We conclude by
	        consider the role future low-energy electron-capture sources could play in this effort.	
	        \end{abstract}
		
		
	\end{frontmatter}
	
	\newpage
	
	\thispagestyle{empty}
	\tableofcontents
	
	%to begin the line numbers: 
	%\linenumbers

	%beginning of the core of the manuscript
	\newpage
	\section{Introduction}\label{sec:intro}
	
The gallium anomaly can be stated:
\textit{The measurements of the charged-current capture rate of neutrinos on \nuc{71}{Ga} from strong radioactive sources have yielded results below those
expected, based on the known strength of the principal transition supplemented by theory.}
The mystery of this anomaly, which has persisted for over two decades, deepened with recent, high-precision results from the Baksan Experiment on Sterile Transitions (BEST) \cite{BESTprl,BESTprc}.

The data that initially gave rise to the anomaly were obtained from calibration tests of two radiochemical detectors, SAGE and GALLEX, that were designed to 
probe low-energy components of the solar neutrino flux.   It was anticipated that the calibration tests would be unusually free of uncertainties.  First, the neutrino sources employed
are well understood, as their intensities can be measured in multiple ways, to sub-1\% precision, and as the spectra they produce are lines with precisely known energies
and branching ratios.  Second, the cross section for neutrino absorption on \nuc{71}Ga is tightly constrained by the known electron-capture lifetime of \nuc{71}Ge, which establishes a
lower bound on the cross section, leaving only a $\sim$6\% correction due to transitions to excited states in \nuc{71}Ge to be determined through a combination of experiment and theory:
the cross section has been recently re-examined in an analysis that carefully propagates all known sources of uncertainty \cite{cross_2023}.  Third, the efficiency of the extraction of Ge from the
Ga targets is independently verified by tracer experiments, in every experimental run.    While questions have been raised about detector operations including the \nuc{71}Ge extraction efficiencies, no plausible experimental explanation for the anomaly has been identified.    Consequently, the discrepancies found in the calibration experiments are concerning, and indeed
have been taken as evidence for sterile neutrinos (\nust).

The gallium anomaly found in the four original calibration experiments represents about a 2.5$\sigma$ deviation from expectations.
With the completion of BEST, the deviation has risen to 6$\sigma$.
This level of significance might not normally 
warrant suggestions of new physics. 
It has happened in this case because the gallium anomaly is one of several that have arisen in low-energy neutrino experiments.
A broad overview of sterile neutrinos and anomalies motivating them
can be found in various community white papers \cite{Acero2022,Adhikari2016,Abazajian2012}.   Yet, despite this supporting evidence as well as the theoretical enthusiasm
for additional neutrino species, no compelling overall sterile-neutrino explanation
for the collection of anomalies has emerged: there is some tension among the sterile neutrino parameters required to account for each anomaly.
Indeed, the freedom one has in introducing sterile neutrinos --
the specific mechanism, their number, and their masses and mixings  -- is considerable, making it difficult to either rule out or confirm such an explanation.  Further, additional
neutrino species can be accompanied by other new neutrino physics.   The  $6\times6$ mixing matrix that arises for three neutrino flavors with both Dirac and Majorna mass terms contains
various mixing angles and phases~\cite{Kobzarev1980} associated with possible CP~\cite{NUNOKAWA2008338} or CPT violation~\cite{Barger_2000}.  Other new physics could include
non-standard neutrino interactions~\cite{Farzan2018}, neutrino decay~\cite{Lindner_2001}, Lorentz violation~\cite{Alan_Kosteleck__2004}, extra dimensions~\cite{Barbieri_2000}, energy dependent mixing parameters~\cite{Babu2022EvaryingParam},
 dark photons~\cite{Alonso_lvarez_2021}, neutrinos coupled to fuzzy dark matter or dark energy~\cite{Brdar2023}, and bulk neutrinos~\cite{Machado2012}.  So despite the expectation that new neutrino species may exist, the flexibility of the theory has made
it difficult to access the plausibility of sterile neutrinos as the explanation for the various anomalies.  For the same reason, it is difficult to
design an experiment to either verify or falsify a hypothesized sterile neutrino.  Specifically,  BEST was not designed for this purpose.  Instead, it was envisioned as a 
high-sensitivity test of the gallium anomaly. While it
increased the significance of the anomaly, it failed to provide more specific evidence of oscillations through a tell-tale 
variation of signal with distance.

In this review, we describe the status of the gallium anomaly, with special emphasis on the most recent results from BEST, which utilized a \nuc{51}Cr 
neutrino source of unprecedented strength.   The resulting increase in the significance of the gallium anomaly is notable, though the results of BEST and 
the four earlier calibrations could still be attributed to an unlikely statistical fluctuation.
We discuss the various cross-checks on the experimental methods that have been made
over the three decades of gallium detector operations.   We conclude by discussing possible steps that could be taken in the future, to resolve this 
perplexing situation.  	
	
\section{History}\label{sec:history}
	
In 1968 the first results from the Homestake chlorine solar neutrino experiment~\cite{Cleveland1998} were announced.  Ray Davis's radiochemical detector made
use of the reaction \nuc{37}{Cl}$(\nu_e,e^-)$\nuc{37}{Ar} to observe \nuc{8}B and \nuc{7}Be solar neutrinos.  This technique was first suggested by Pontecorvo~\cite{Pontecorvo1946}, then explored in more detail by Alvarez \cite{Alvarez1949}, who was interested in doing a reactor experiment to test
whether neutrinos were Majorana particles.   The Davis detector consisted of 615 tons of perchloroethylene (C$_2$Cl$_4$), placed inside a steel containment vessel 
that had been constructed on the 4850-ft level of the Homestake gold mine.  As a noble gas that does not interact chemically, argon can be extracted with 
high efficiency ($\sim$95\%) from large volumes of organic liquid. The $\sim$35-d half-life of \nuc{37}Ar  is nearly ideal, allowing tank
concentrations to build up over a saturation time of about two months, yet permitting
\nuc{37}Ar counting via electron capture (EC).  On two-month intervals, approximately ten \nuc{37}{Ar} atoms produced by solar neutrino reactions would be extracted from the volume and counted in small proportional counters, which recorded the emitted x rays and Auger electrons produced after \nuc{37}Ar EC, as the
K-shell vacancy is filled.  Measurements continued until 2002, when the Homestake mine was closed.  The end result was a neutrino capture rate of 2.56$\pm$0.16$\pm$0.16~SNU\footnote{1 SNU (solar neutrino unit) = \cpowten{1}{-36}\ /(target atom - s).}, about one-third that predicted by the standard solar model (SSM)
~\cite{Magg2022,Serenelli2011,Sackman1990,Turck1993,Bahcall1995}.

Approximately 75\% of the events in the chlorine detector came from the capture of more energetic \nuc{8}B neutrinos.  In the SSM the flux of these neutrinos
varies as $\phi(^8$B)$\sim \mathrm{T}_c^{22}$, where $\mathrm{T}_c$ is the solar core temperature.  This prompted many early solutions
of the so-called ``solar neutrino problem" in which the SSM was modified in ways that would reduce the core temperature by $\sim 5$\%, thereby
eliminating the discrepancy between observation and theory.  Others proposed solutions invoked new weak-interaction physics, including
neutrino oscillations and neutrino decay, or questioned whether there might be a hidden flaw in the experiment, as the radiochemical method
is indirect.  An account of the many inventive solutions can be found in the entertaining review of Ref. \cite{account}. 

The chlorine detector was ahead of its time:  two decades passed before others could build detectors to cross check the Homestake result.  In the early
1980s the proton decay experiment Kamiokande I was re-instrumented to detect lower energy events, which enabled Kamiokande II/III to measure
the high-energy portion of the \nuc{8}B solar neutrino flux via the reaction $\nu+e \rightarrow \nu^\prime +e^\prime$.  The Cerenkov light produced by the
recoiling electron was observed in the three-kiloton water detector.  Kamiokande II operated with a 9 MeV threshold for the electron energy, which was
lowered to 7 MeV in Kamiokande III \cite{Fukuda1996}.  The first results from Kamiokande II were announced in 1989.  The neutrino event  rate was
approximately 50\% that expected, based on the SSM \nuc{8}B flux prediction.  The fact that Kamiokande measured neutrinos event-by-event, provided some
information on the shape of the neutrino spectrum, and 
largely confirmed the results of the Homestake experiment had significant impact.

\begin{figure}
  \centering
  \includegraphics[width=0.7\columnwidth]{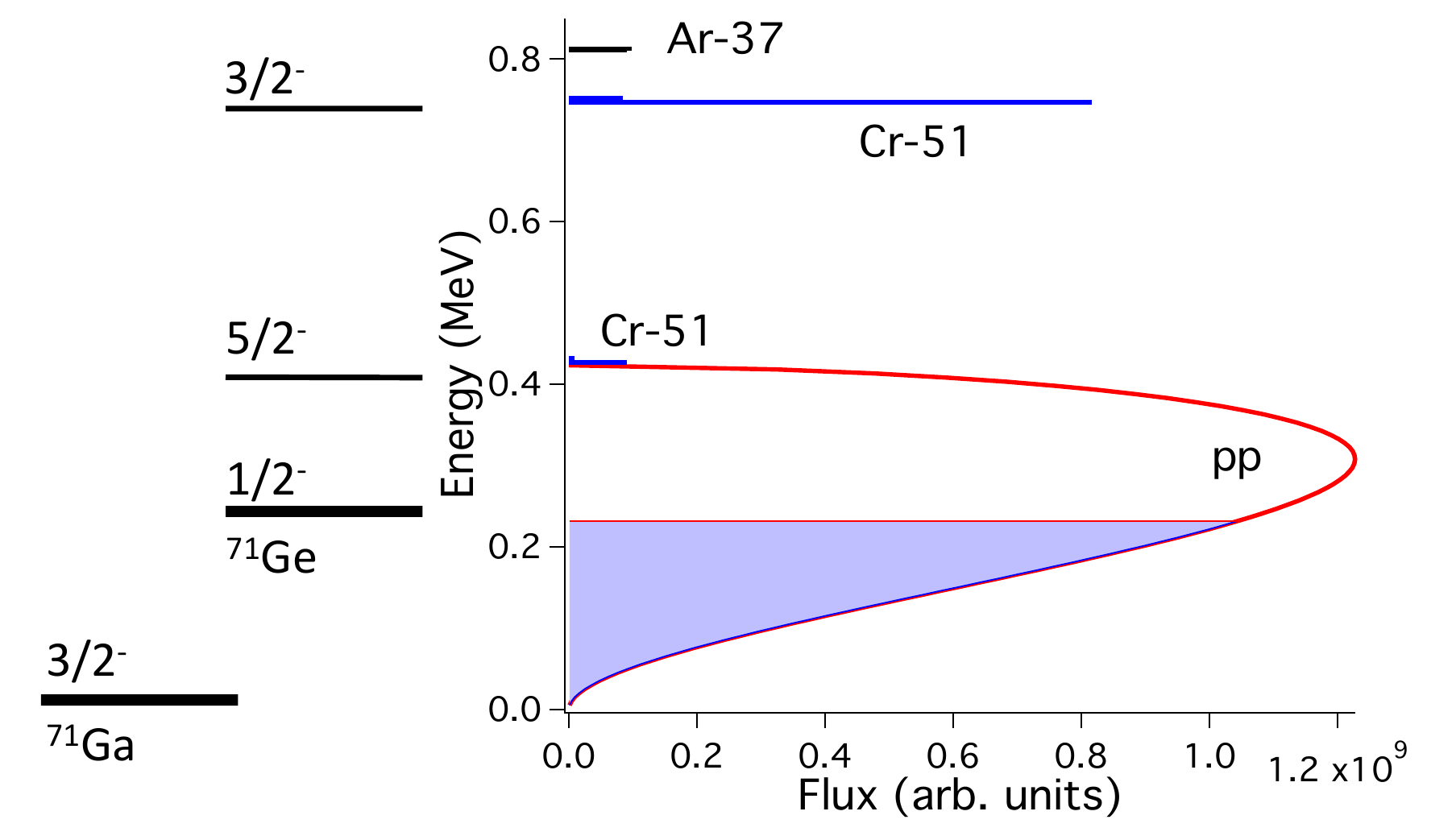}
  \caption{The \nuc{71}{Ga}-\nuc{71}{Ge} level diagram compared to the neutrino spectra from the pp fusion (red curve), \nuc{51}{Cr} (lower blue bars), and \nuc{37}{Ar} (top black bar). The shaded region denotes the portion of the pp spectrum below threshold for neutrino capture on \nuc{71}{Ga}. }
  \label{fig:LevelDiagram}
\end{figure}

Kuzmin~\cite{Kuzmin1965} had proposed using \nuc{71}{Ga} as the target for a radiochemical solar neutrino experiment
due to the 234~keV threshold of the reaction \CaptReac\ (see Fig.~\ref{fig:LevelDiagram}) and 11.43 d half-life of \nuc{71}Ge.  
The low threshold provides sensitivity to
the low-energy pp neutrinos --  those generated in the first step of the pp chain via proton-proton (pp) fusion -- which are produced in a 
$\beta$-decay spectrum with an endpoint of 420 keV.  The SSM flux of pp neutrinos
is about four orders of magnitude higher than that of the \nuc{8}B neutrinos.
Further, in contrast to the temperature-dependent \nuc{8}B neutrinos, the flux of the pp neutrinos is 
constrained by the Sun's luminosity, assuming only a steady-state Sun and standard-model
weak interaction physics.   With these assumptions, a minimum counting rate of 79 SNU was predicted for this detector.  Consequently,
a rate lower than this bound would point to new neutrino physics, an exciting result.

In the 1970s work began on two possible approaches to the chemistry of this detector, 
one employing gallium as a GaCl$_3$ solution and the other as a metal. In the former,
after an exposure of about three weeks, the produced germanium was recovered
as GeCl$_4$ by bubbling nitrogen through the solution, then scrubbing the gas. The
Ge was further concentrated and purified, converted into GeH$_4$, then counted
in miniaturized gas proportional counters similar to those used in the chlorine
experiment.  This procedure was employed in the GALLEX/GNO experiment.

The SAGE experiment exploited the fact that gallium metal is a liquid at slightly above room temperature.
The produced \nuc{71}Ge is 
separated from the metal by mixing into the gallium a solution of hydrogen peroxide and dilute
hydrochloric acid, which produces an emulsion, with the germanium migrating to
the surface of the emulsion droplets where it is oxidized and dissolved by hydrochloric
acid. The Ge is extracted as GeCl$_4$, purified and concentrated, synthesized
into GeH$_4$, then counted as in the GALLEX experiment. In both GALLEX and
SAGE, the overall efficiency of the chemical procedures can be determined by
introducing Ge carrier.

\begin{figure}
  \centering
  \includegraphics[width=0.4\columnwidth]{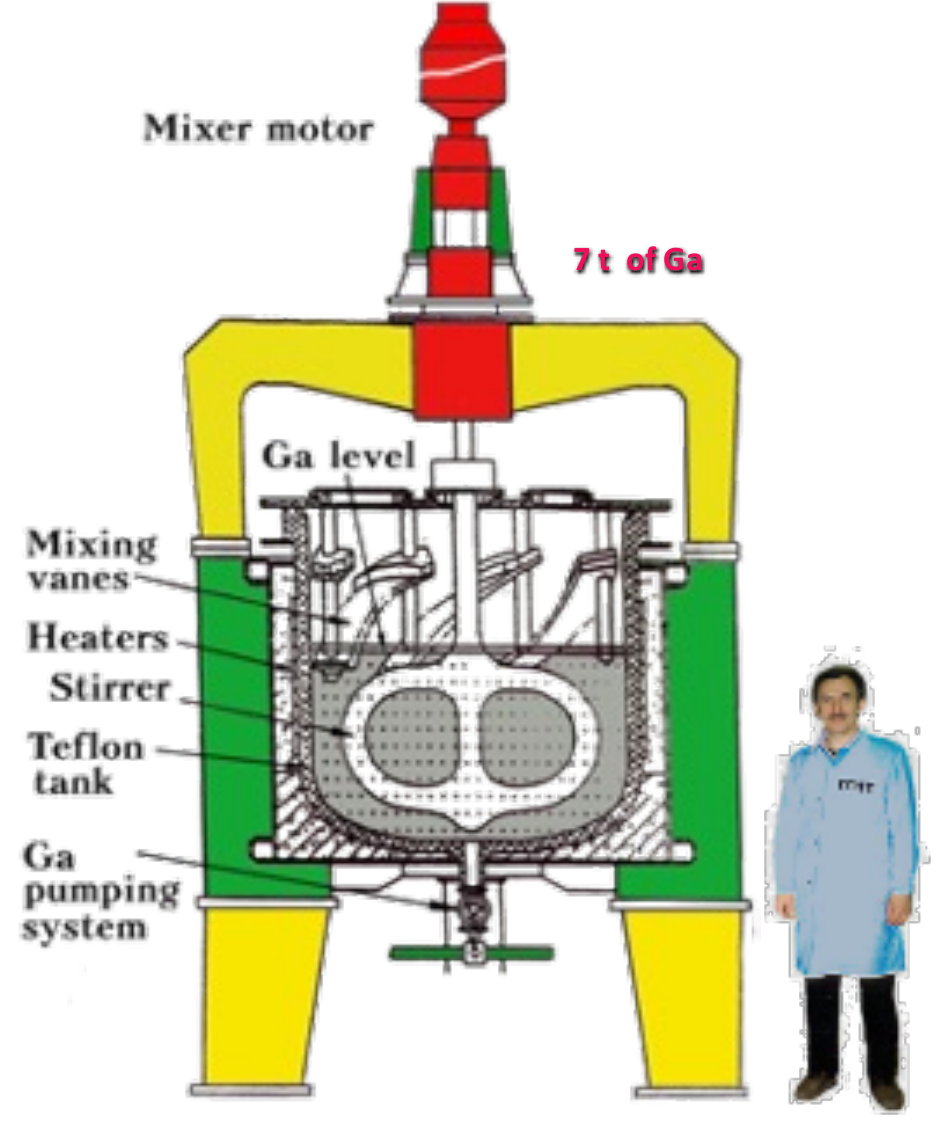}
  \includegraphics[width=0.4\columnwidth]{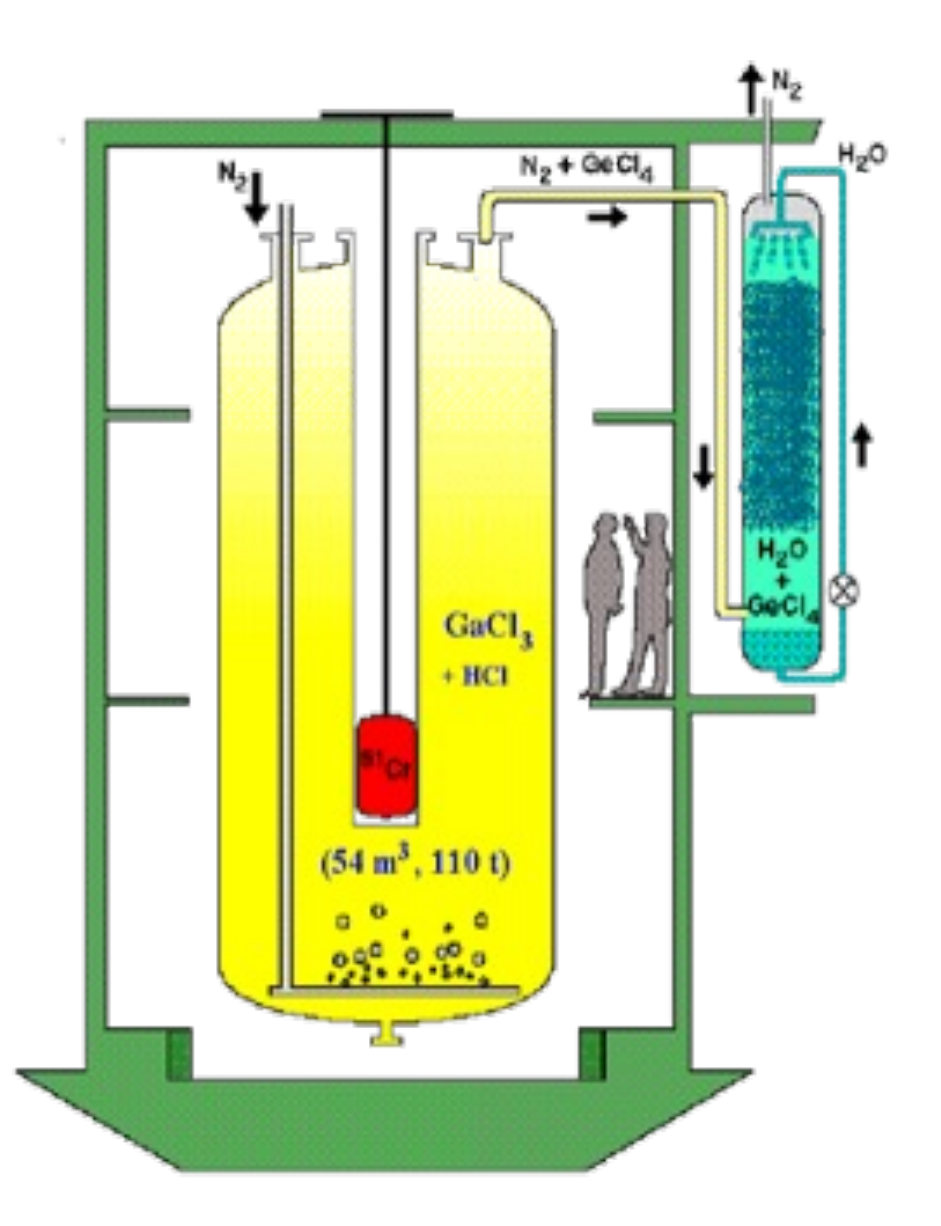}
  \caption{Left: Sketch of one of the SAGE reactors used for solar neutrino measurements. Right: Sketch of the GALLEX neutrino source experimental layout. (Figure from https://www.mpi-hd.mpg.de/lin/images/tank.gif) }
  \label{fig:ExperimentalLayouts}
\end{figure}

 The SAGE and GALLEX/GNO experiments (Fig.~\ref{fig:ExperimentalLayouts}) began in the late 1980's with operations stretching into the new century. The deduced counting rates (about 70~ SNU~\cite{abdurashitov2009measurement,ALTMANN2005174}) were again low compared to the SSM prediction (137~SNU~\cite{Bahcall1995}).  They were also just below the minimum astronomical value,
 suggesting that the solar neutrino problem extended beyond \nuc{8}B neutrinos, affecting also the lower energy portion of the solar neutrino 
 flux.  Perhaps most important, the combination of the Cl, Kamiokande, and SAGE/GALLEX results indicated a pattern of solar neutrino fluxes 
 incompatible with any choice of the solar core temperature T$_c$.  Consequently, a new solution involving new particle physics became
 plausible, giving impetus to three new experiments,
 Borexino, Super-Kamiokande, and the Sudbury Neutrino Observatory, that were to vastly improve our knowledge of both the Sun and the basic
 physics of neutrinos.    Super-Kamiokande~\cite{Abe2016} measurements of both atmospheric and solar neutrinos and SNO~\cite{PhysRevC.88.025501} 
 measurements separating electron from heavy-flavor solar neutrinos showed that neutrinos are massive and undergo flavor oscillations.
 Borexino's patient campaign to map out the entire spectrum of solar neutrinos revealed the transition between vacuum-dominated and
 matter-dominated solar neutrino oscillations, thereby providing the first information on the ordering of the three light neutrino mass eigenstates.
 There are several excellent reviews summarizing the history of the solar neutrino problem and its resolution (see ~\cite{Nakahata2022,OGZBS2021,HRS2013}). 
 
 This review focuses on follow-up calibration measurements done by the SAGE and GALLEX/GNO experiments that
 generated new questions, still not resolved.
 The critical role of gallium experiments in underscoring the seriousness of the solar neutrino problem, combined with the recognition that the chemistry
of these detectors was more complicated than that of the chlorine detector, led to proposals to make end-to-end cross checks of operations,
by exposing the detectors to well-calibrated artificial neutrino sources.
In the mid 1990's and early 2000's, four high-activity, artificial-source experiments were conducted, three with \nuc{51}{Cr} and one with \nuc{37}{Ar}.
These electron-capture line sources produce low energy neutrinos, similar to the \nuc{7}Be solar neutrino line source.
As the source intensities were typically on the order of one MCi, the counting rates the sources induced in the detectors were high, by solar neutrino
counting-rate standards. The intensities of the sources were calibrated by several means, and very well established.
Yet in combination they showed a rate 13\% below that expected, albeit with a $\pm$5\% uncertainty. 

By the time Super-Kamiokande and SNO had produced their results, other nagging neutrino-physics discrepancies had been pointed out.  One of the
earliest and perhaps most widely discussed came from the LSND experiment~\cite{Athanassopoulos_1996}, which searched for $\bar{\nu}_\mu \rightarrow \bar{\nu}_e$ using $\bar{\nu}_\mu$s from the
decay of muons at rest and $\nu_\mu \rightarrow \nu_e$ using $\nu_\mu$s from $\pi^+$s decaying in flight, observing events in a liquid scintillator detector.  Difficulties the collaboration encountered in understand the spectrum of events led them to suggest oscillations
into sterile neutrino states \cite{Aguilar2001}.  These and other similar claims, reviewed in \cite{Acero2022,Adhikari2016,Abazajian2012}, provided additional motivation for BEST -- a fifth gallium calibration 
experiment employing a source of unprecedented intensity, performed with the existing SAGE detector, though redesigned
(see Fig. \ref{fig:ExperimentalLayouts}) to have some sensitivity to 
oscillation baselines.  The results from that experiment, as well as those from the four earlier gallium neutrino source experiments, form the
focus of this report. 
	
\subsection{Gallium Radiochemical Radioactive Source Measurements}
Exposing the gallium to a known source of $\nu_e$s and then carrying out the routine procedures of \nuc{71}Ge extraction and
counting provides an end-to-end cross-check of all experimental procedures, including the \nuc{71}Ga($\nu_e,e^-$)\nuc{71}Ge
cross section.   The efficiencies of various extraction and counting steps during solar neutrino operations were already calibrated 
through auxiliary measurements, providing a systematic verification of experimental performance of both GALLEX and SAGE.
Thus one would not have expected a
different result in a neutrino calibration experiment.  Indeed, because carrier experiments verify that the extraction of Ge is highly efficient,
it had been argued that a high-statistics neutrino source experiment could be viewed
as a measurement of the neutrino absorption cross section \cite{Hata1995,Haxton1998}. Though tightly constrained by experiment, the cross section does have some
residual dependence on theory due to transitions to two excited states in \nuc{71}Ge.  Yet we will see that the BEST/gallium anomalies
cannot be attributed solely, or even primarily, to this theory uncertainty.
~\\

\begin{table}[htp]
\caption{Line neutrinos from the source isotopes, taking into account the probabilities for K, L, and M capture ~\cite{TabRad_v7,TabRad_v1,TabRad_v0}.}
\begin{center}
\begin{tabular}{lccc}
\hline
Isotope 					&	$\tau_{1/2}$ (d)		&$E_{\nu}$ (keV)	& $f_{E_{\nu}}$	(\%)	\\
\hline\hline
\multirow{3}{*}{\nuc{37}{Ar}} 	& \multirow{3}{*}{35.0} 	&813.8			&\phantom01.11$\pm$0.01	\\
                                                  &                                        &813.5               &\phantom08.66$\pm$0.01 \\ 
						&					&810.7			&90.23$\pm$0.01	\\
\hline
\multirow{6}{*}{\nuc{51}{Cr}} 	& \multirow{6}{*}{27.7} 	&752.4			&\phantom01.40$\pm$0.01	\\
                                                  &                                        &751.8                       &\phantom08.42$\pm$0.01   \\
						&					&746.5			&80.25$\pm$0.01	\\
						&					&432.3			&\phantom00.15$\pm$0.01	\\
						&                                        &431.7                      &\phantom00.92$\pm$0.01  \\
						&					&426.4			&\phantom08.86$\pm$0.01	\\
\hline\hline
\end{tabular}
\end{center}
\label{tab:SourceIsotopes}
\end{table}%

The electron-capture lines produced by \nuc{37}{Ar} and \nuc{51}{Cr} sources are given in Table~\ref{tab:SourceIsotopes},
taking into account K, L, and M capture and, in the case of \nuc{51}{Cr}, the $\sim$10\% probability of capture to the first 
excited state of \nuc{51}V.
The \nuc{51}Cr and \nuc{37}Ar sources can be produced by irradiating isotopically enriched \nuc{50}{Cr} (as the natural abundance of
this isotope is just 4.3\%) or natural Ca or CaO targets in a high-flux reactor, making use of
the $(n,\gamma)$ and $(n,\alpha)$ reactions, respectively.  

As \nuc{71}{Ge} has an 11.43-d half-life~\cite{hampel1985}, the exposures for \nuc{71}Ga($\nu_e,e^-$) were typically 5-10~days  (5 d for GALLEX, and 10 d for BEST). The produced Ge was then extracted, along with Ge carrier, and a counter gas synthesized. This gas was inserted into a small proportional counter and counted for a few months. The important nuclear and atomic physics data for \nuc{71}{Ge} decay are given in Table~\ref{tab:GeDecay}.  The source experiments are described in the following subsections.
~~\\

\begin{table}[htp]
\caption{Key features of the EC decay of \nuc{71}{Ge}~\cite{Genze1971,Neumann1982}.}
\begin{center}
\begin{tabular}{lcl}
\hline
Shell Capture 				&	$f_c$					& Emissions		\\
\hline\hline
\multirow{3}{*}{K} 	& \multirow{3}{*}{88\%} 			&41.5\% 10.367-keV Auger e$^-$	\\
						&							&41.2\% 1.2-keV Auger e$^-$ \& 9.2-keV x ray	\\
						&							&5.3\% 0.12-keV Auger e$^-$ \& 10.26-keV x ray	\\
\hline
L				 	&	10.3\%					&1.2-keV Auger e$^-$		\\
\hline
M 				 	&	1.7\%					&0.12-keV Auger e$^-$		\\
\hline\hline
\end{tabular}
\end{center}
\label{tab:GeDecay}
\end{table}%

\subsection{SAGE \nuc{51}{Cr} and \nuc{37}{Ar} Source Measurements}
\label{sec:SAGE}
\begin{figure}
  \centering
  \includegraphics[width=0.6\columnwidth]{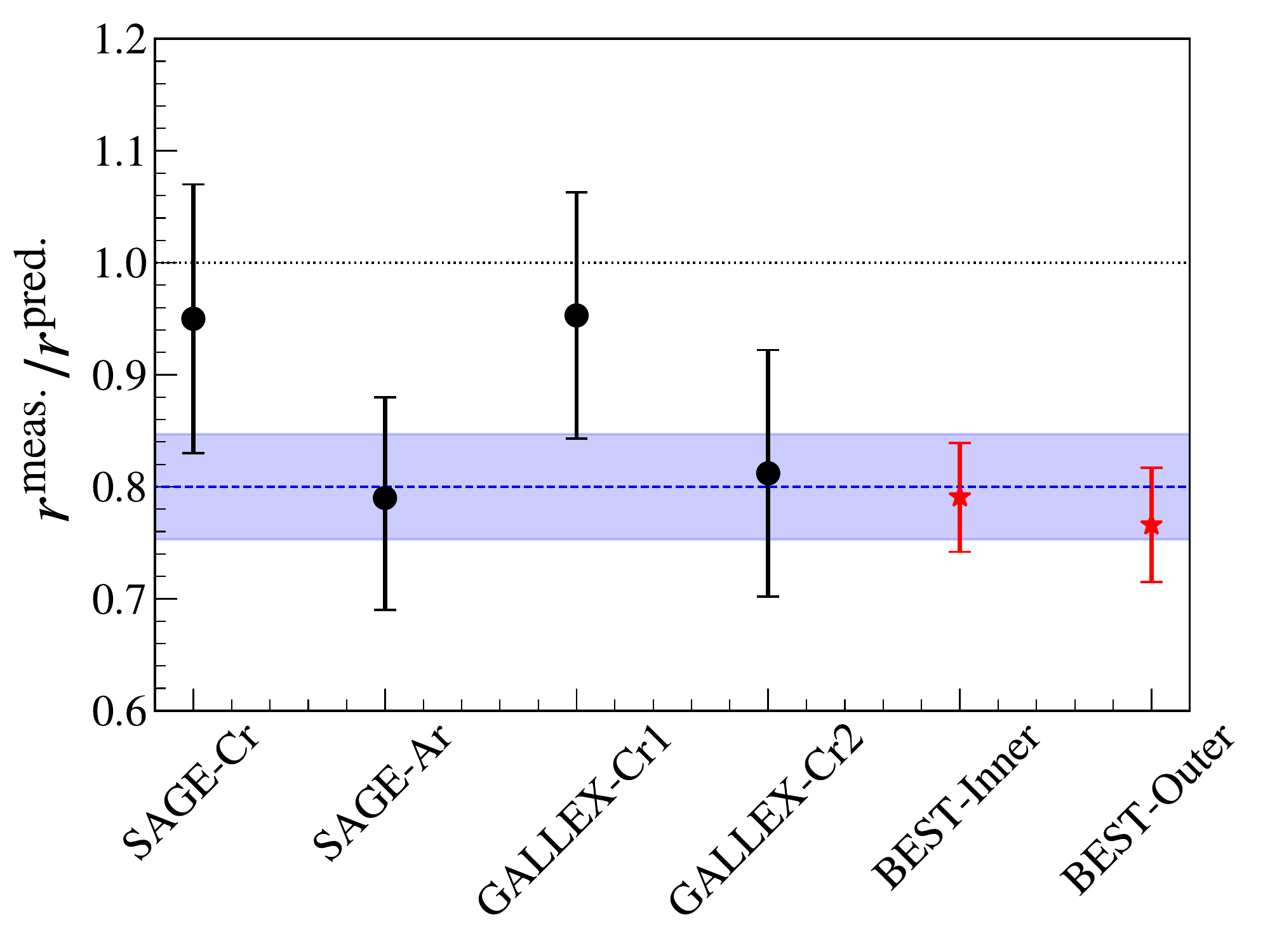}
  \caption{The ratio of the measured \nuc{71}{Ge} production rate to the predicted rate for all 6 measurements. The dotted blue line (shading) is the best fit (uncertainty) to all 6 results.}
  \label{fig:ProdRatio}
\end{figure}

The SAGE source experiments followed the same procedures used in SAGE solar neutrino measurements, with one important difference. During
solar neutrino operations each reactor contained a stirring mechanism, installed to evenly disperse throughout the Ga target the small quantity of natural Ge carrier that was added.  This mechanism took up a great deal of space and would have interfered with source installation. Hence a reactor without that mechanism was employed for the source exposures. This special reactor held 13~t of Ga compared to the 7~t it would have held had the stirring mechanism remained in place. However, lacking a stirring mechanism, this reactor could not be used for the extraction chemistry.  Instead,
after exposure, the Ga was pumped to other reactors for this step. After Ge extraction, the procedures followed those used for decades in solar neutrino running. The results of the \nuc{51}{Cr} and \nuc{37}{Ar} source measurements are given in Table~\ref{tab:Activities} and depicted in Fig.~\ref{fig:ProdRatio}.
		 
The GALLEX, SAGE, and BEST experiments followed very similar procedures in fabricating intense \nuc{51}{Cr} sources, as summarized in Sec.~\ref{sec:CrSources}.  In addition, SAGE performed one calibration with an \nuc{37}{Ar} source, which produces a neutrino line very similar in energy
to the solar \nuc{7}Be line.   That source was produced by irradiating CaO (12.36~kg Ca) in the fast neutron breeder reactor BN-600 at Zarechny, Russia. The CaO was dissolved in a nitric acid solution and the \nuc{37}{Ar} was extracted by a He purge~\cite{sage2006argon}. The source activity of the Ar was estimated by six distinct procedures. First, the volume and isotopic composition of produced gas was measured. Second, the mass of Ar introduced into the source container was measured. Third and fourth, the heat output of the source was measured by calorimetry at Zarechny before shipping to Baksan and after arrival using the same apparatus as for the SAGE Cr measurements, noting that the energy released by \nuc{37}{Ar} is $2.751\pm0.021$~keV/decay\cite{sage2006argon}.  Fifth, after exposures of the Ga were completed, the source was returned to Zarechny and the remaining \nuc{37}{Ar} activity of samples was measured.  Lastly, these final samples were analyzed for isotope dilution. The final value for the activity (409$\pm$2~kCi), determined from the weighted average, has an uncertainty of 0.5\%.
	
\subsection{GALLEX \nuc{51}{Cr} Source Measurements}
\label{sec:GALLEX}
	The GALLEX tank was filled with 101 tons GaCl$_3$, acidified to 2~M in HCL.  The target contained 30.3~t of natural Ga (12~t \nuc{71}{Ga}). Ge forms the volatile molecule GeCl$_4$ which can be extracted from the non-volatile GaCl$_3$ by bubbling an inert gas through the solution~\cite{Anselmann1992}.  The GALLEX synthesis of GeH$_4$ follows procedures similar to those employed by SAGE  and BEST, described in Sec.~\ref{sec:BESTOverview}. The counting and analysis procedures used by GALLEX and SAGE are also very similar. The GALLEX experimental setup is depicted in Fig.~\ref{fig:ExperimentalLayouts}.
	
The GALLEX source had a 50-cm diameter, whereas the SAGE source had a diameter of only 8~cm. Furthermore the GaCl$_3$ solution in GALLEX has a lower \nuc{71}{Ga} atomic density than the metallic Ga used by SAGE.  Consequently, though the GALLEX sources were a factor three stronger than those used in SAGE, the Ge production rates and hence the precision of the calibrations were similar in the two experiments.
	
The GALLEX solar neutrino results were reanalyzed~\cite{gallex2010reanalysis} after all solar neutrino runs were completed in 2003. The counter efficiencies could then be measured with high statistics using internal radioactive sources that would have compromised their performance for low-background measurements. In addition, advanced pulse shape analysis~\cite{ALTMANN1996} was implemented to better distinguish signal from background.  The results of the two \nuc{51}{Cr} measurements are given in Table~\ref{tab:Activities} and shown in Fig.~\ref{fig:ProdRatio}.

\subsection{Results from the Early Source Measurements}
\label{sec:D}
A summary of the GALLEX, SAGE and BEST results are given in Table~\ref{tab:Activities}. It should be noted that GALLEX updated their analysis and efficiencies~\cite{ALTMANN2005174,gallex2010reanalysis}. The SAGE reference~\cite{sage2006argon} quoted different values for the measured-to-predicted ratios than those that were eventually established based on private communications regarding the upcoming efficiency updates. The differences are modest compared to the $\sim$10\% uncertainties (1.01 instead of 1.00, and 0.84 instead of 0.81).
	
The typical precision achieved in each GALLEX and SAGE source experiment was $\gtrsim$ 10\%.  When combined, the four experiments yield a ratio of
observed to expected counts of $R$=0.87$\pm$0.05~\cite{abdurashitov2009measurement}, a deviation of $\sim$ 2.5$\sigma$ from 1.  Although not statistically convincing, the discrepancy generated a great deal of speculation as to possible causes, and became known as the ``Ga anomaly."  
In particular, the discrepancy has been often cited as tentative evidence for sterile neutrinos and  \nuel\ $\rightarrow$ \nust\ oscillations \cite{Acero2022,Adhikari2016,Abazajian2012}.

If one assumes a two-component oscillation into a sterile state, the survival probability for a neutrino of energy $E_{\nu}$ detected a distance $L$ from its source is 
\begin{equation}\label{eqn:OscilationProbability}
P_{ee}(E_\nu,L) =1 - \mbox{sin}^22\theta ~\mbox{sin}^2 \left({\pi L \over L_\mathrm{osc}} \right)~~~\mathrm{with}~~~{\pi \over \left[ L_\mathrm{osc}/\mathrm{m}\right]} = 1.27 \,  {\left[\Delta m^2/\mathrm{eV}^2\right] \over \left[E_{\nu}/\mathrm{MeV}\right]}
\end{equation}
where $\Delta m^2= m_2^2-m_1^2$, $m_i$ is the mass of the $i$th mass eigenstate, and $\theta$ is the mixing angfe.  \

To determine
the impact of oscillations on the detection rate in the Ga calibration and BEST experiments, one must take into account the complex geometry of the detector and the source.  
Denoting the target and source volumes by $V_d$ and $V_s$, respectively, the neutrino capture rate $r$ is
\begin{equation} 
r= \sum_{i=1}^6 f_i ~\int_{V_d}  d \vec{x}_d ~n(\vec{x}_d) ~\int_{V_s}  d \vec{x}_s ~a(\vec{x}_s) ~{1 \over |\vec{x}_d-\vec{x}_s|^2}~\sigma(E_\nu^i)~P_{ee}(E_\nu^i, |\vec{x}_d-\vec{x}_s|)
\end{equation}
where $i$ indexes the neutrino lines produced in \nuc{51}Cr EC (the six lines listed in Table \ref{tab:SourceIsotopes}), $f_i$ is the fraction of the neutrinos emitted in line $i$,
$E_\nu^i$ is the line energy, $\sigma(E_\nu^i)$ is the \nuc{71}Ga neutrino absorption cross section at that energy, $n$ is the target \nuc{71}{Ga} number density, 
$a$ is the source activity density,  and $P_{ee}(E_\nu^i,D)$ is the oscillation survival probability for that line at a distance
$D \equiv |\vec{x}_d-\vec{x}_s|$ from the source. Under the assumption that the target number and source activity densities are uniform, this can be rewritten as
\begin{equation}
r= \sum_{i=1}^6 f_i ~\sigma(E_\nu^i)\,N\,A~\int  dD ~{1 \over D^2}~P_{ee}(E_\nu^i, D)~P(D) 
\label{eq:first}
\end{equation}
where $N$ is the number of \nuc{71}Ga nuclei in the target, $A$ is the total source activity, and 
\begin{equation}
P(D) \equiv {1 \over V_d}\int_{V_d}  d \vec{x}_d  ~~{1 \over V_s}\int_{V_s} d \vec{x}_s ~\delta(D- |\vec{x}_d-\vec{x}_s|)~~~\mathrm{~with~} ~~~\int  dD ~P(D) =1.
\label{eq:second}
\end{equation}

Together, Eqs (\ref{eq:first}) and (\ref{eq:second}) factor the cross section and oscillation physics from the target and source geometry, encoding the latter in a probability distribution $P(D)$ describing the likelihood that a given neutrino
interacting in the target did so after traveling a distance $D$ from the point where it was produced.  This quantity is computed by Monte Carlo integration due to the complexity of target and detector geometry, after which it can 
be used in detailed oscillation studies.  

One can also define an effective path length in the detector
\begin{equation} 
\langle L \rangle \equiv {1 \over 4 \pi}  \int_{V_d}  d \vec{x}_d  ~~{1 \over V_s}\int_{V_s} d \vec{x}_s~ {1 \over |\vec{x}_d-\vec{x}_s|^2}~ \rightarrow~{1 \over 4 \pi} \int_{V_d}  d \vec{x}_d  ~ {1 \over |\vec{x}_d|^2}
\label{eq:third}
\end{equation}
where on the right we have taken the limit where the source radius is much smaller than the distance to the detecting region, so that the source can be approximated as a point.
One can see that $\langle L \rangle$ is basically the average thickness of the detection region: in a detector like BEST with inner and outer regions, if each has approximately the same $\langle L \rangle$,
the detection rate in each volume with be approximately equal, in the absence of oscillations.  The sensitivity to oscillation lengths is also governed by $\langle L \rangle$.
If $L_\mathrm{osc} >> \langle L \rangle$, the detector event rate will not be affected by oscillations;  if $L_\mathrm{osc} << \langle L \rangle$ the detection rate will be reduced by the factor
$1 - {\mbox{sin}^22\theta \over 2}$, but the rapidity of the oscillations will make it impossible to see variations as a function of distance.   However, if $L \sim L_\mathrm{osc}$ and a detector records the positions of events,
variations can be seen.

We relate $\langle L \rangle$ to a weighted distribution involving $P(D)$ by inserting $1=\int dD \, \delta(D-|\vec{x}_d -\vec{x}_s|)$ in Eq. (\ref{eq:third}), yielding
\begin{equation}  
\langle L \rangle = {V_d \over 4 \pi} \int dD \, {1 \over D^2} P(D) \equiv \int dD \, L(D)
\end{equation}
In a Monte Carlo evaluation of Eq. (\ref{eq:third}), if events are binned according the distance D between the neutrino production point in the source and
absorption point in the target, one has in the resulting dimensionless distribution $L(D)$ the information needed to simulate oscillations for any choice of oscillation parameters.
$L(D)$ has the normalization
\begin{equation}
 \int dD \, D^2  L(D) = {V_d \over 4 \pi}
\end{equation}
One can then re-express the rate in Eq. (\ref{eq:first}) in terms of $L(D)$, 
\begin{equation}
r= \sum_{i=1}^6 f_i ~\sigma(E_\nu^i) \,4 \pi \,n \,A~\int  dD ~P_{ee}(E_\nu^i, D)~L(D) 
\label{eq:firstp}
\end{equation}
making it clear that $L(D)$ is the weight one must fold with the oscillation probability to get the rate $r$.

If one interprets the $\sim$ 2.5$\sigma$ discrepancy that emerged from the four SAGE and GALLEX calibration experiments as evidence for a sterile neutrino, the best fit to oscillation parameters yields~\cite{Abazajian2012} $\Delta m^2 \sim$ 2.15 (2.24)~eV$^2$ and sin$^22\theta \sim$ 0.24 (0.5), taking the neutrino absorption cross section from  Ref.~\cite{Bahcall1997} (Ref.~\cite{Haxton1998}).  However, 
the allowed region is very large and the minimum in $\chi^2$ quite flat.  When this flatness is taken into account, it was found~\cite{Abazajian2012} that sin$^2\theta \gtrsim 0.07$ and $\Delta m^2 \gtrsim 0.35$ eV$^2$ at 95\% C.L.,
using the cross section of \cite{Haxton1998}.
There is modest tension between these oscillation parameters and those found from some other
experimental anomalies.  For example, LSND analyses allowed large $\Delta m^2$ consistent with the Ga results, but only if correlated with smaller sin$^22\theta$.  As noted in \cite{Abazajian2012}, however, one could nicely account for both the gallium and reactor neutrino
anomalies by postulating an oscillation into a sterile fourth neutrino.   These inconclusive comparisons with other experiments provided additional motivation for a higher statistics neutrino source experiment, stimulating work on BEST.

\section{The BEST Experiment}\label{sec:BESTOverview}
\label{sec:BEST}
	
\begin{figure}
  \centering
  \includegraphics[width=0.9\columnwidth]{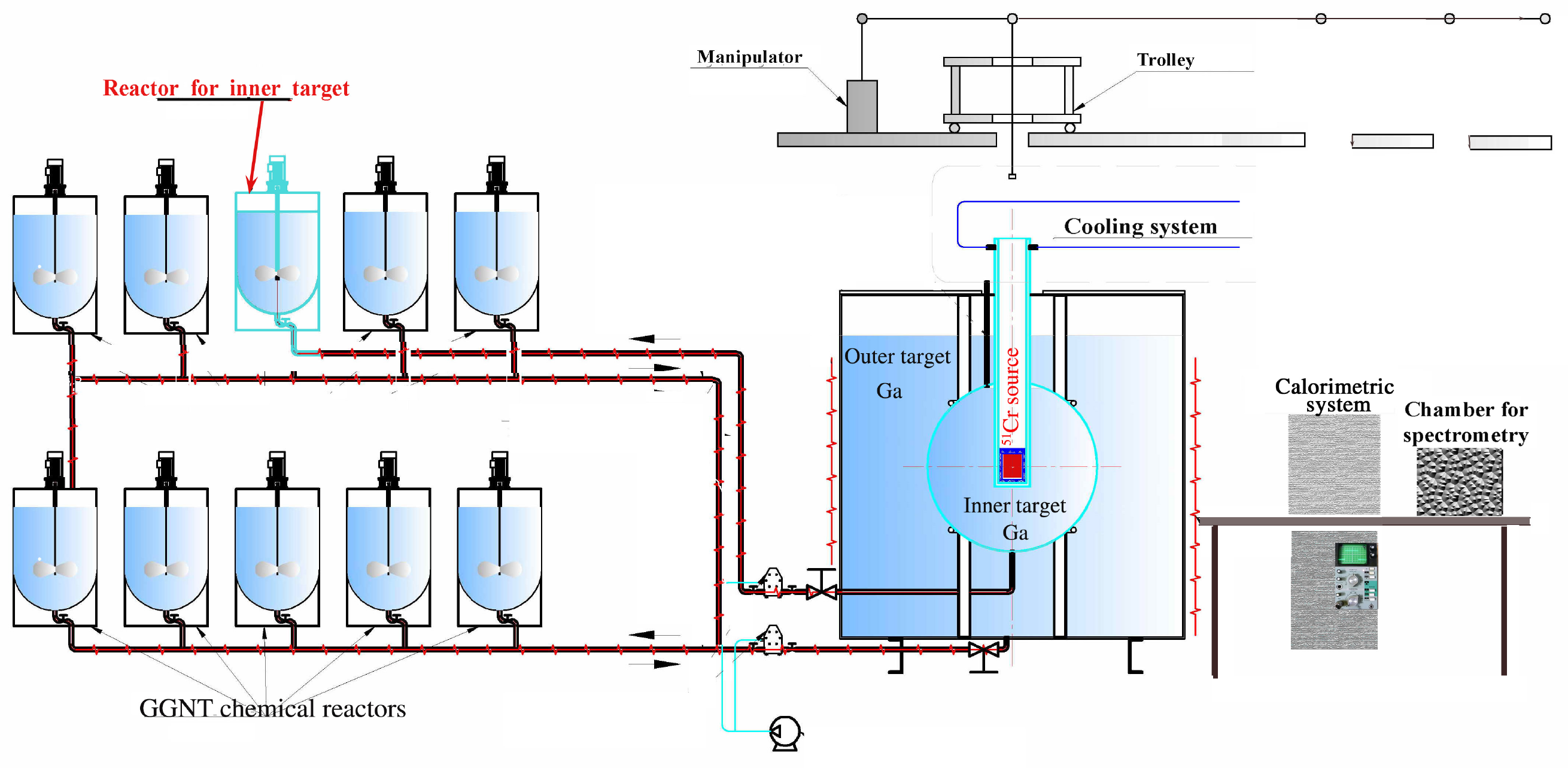}
  \caption{A Cartoon of the BEST experiment configuration.}
  \label{fig:BESTlayout}
\end{figure}

\begin{figure}
  \centering
  \includegraphics[width=0.5\columnwidth]{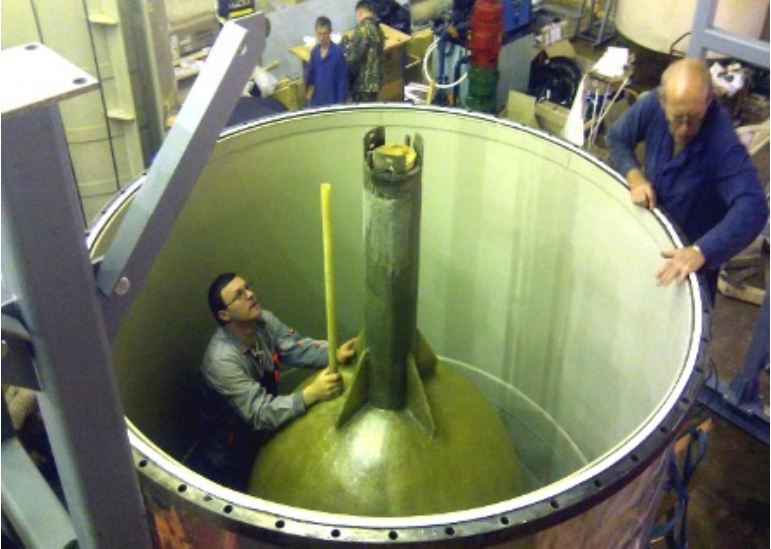}
  \caption{A photograph of the two-target volume during assembly.}
  \label{fig:BESTTarget}
\end{figure}

\begin{figure}
  \centering
  \includegraphics[width=0.3\columnwidth]{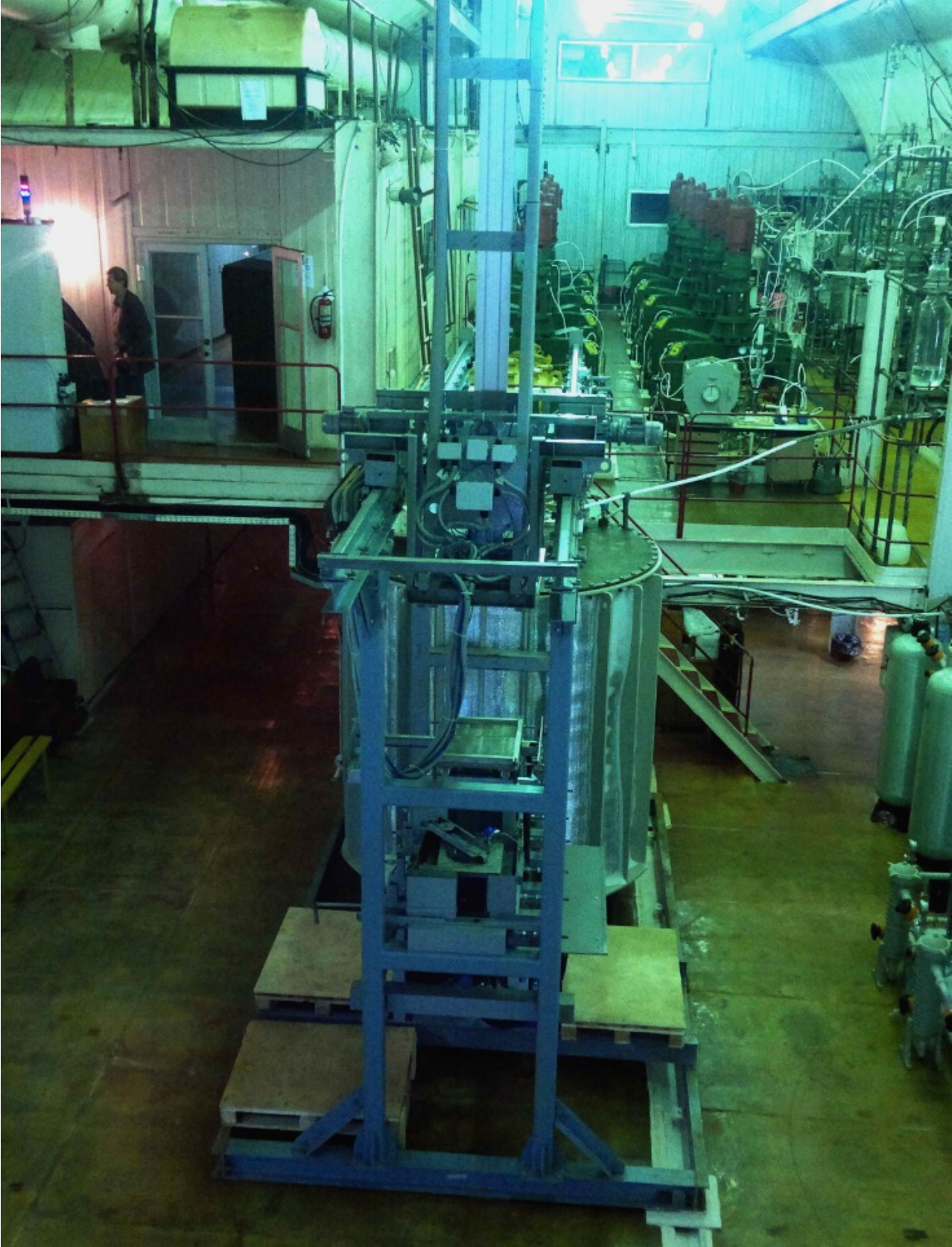}
   \includegraphics[width=0.6\columnwidth]{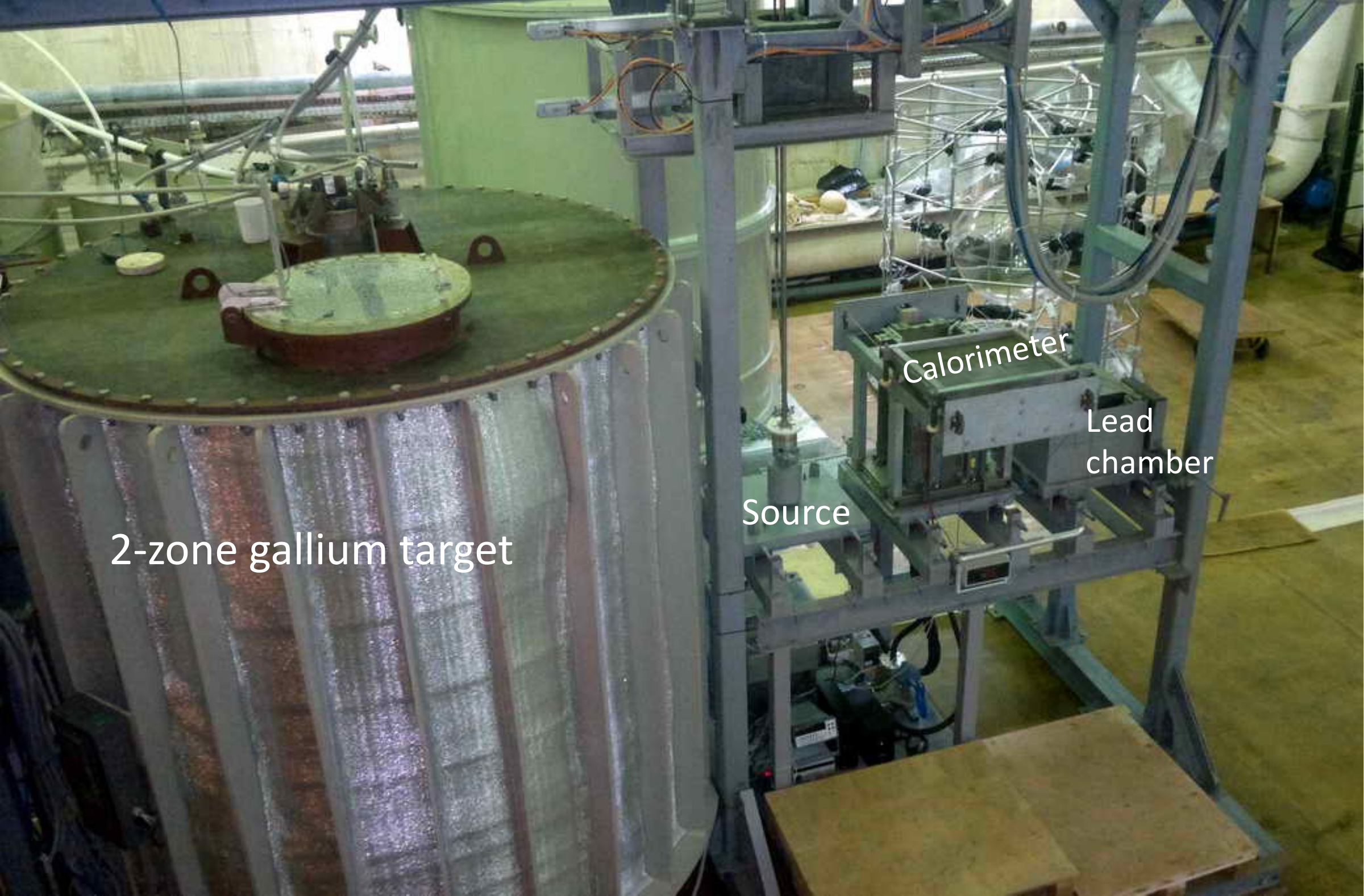}
 \caption{Left: A photograph of the Baksan Neutrino Observatory showing the BEST two-zone target in the foreground and the mixing reactors, with their red cap motor drives in the background. Right: A photograph of the two-zone target, the source, the calorimeter and the lead shield for $\gamma$ spectrum measurements.}
  \label{fig:BESTLabPhotos}
\end{figure}

BEST goals included a higher intensity source, to improve counting statistics, and the introduction of two nested target volumes, so that if oscillations were occurring, variations in the flux with distance might be identified.  Figure~\ref{fig:BESTlayout} shows the experimental layout. Figure~\ref{fig:BESTTarget} shows the nested target volume during construction and Fig.~\ref{fig:BESTLabPhotos} shows photos of the BEST lab at the Baksan Neutrino Observatory. The BEST source, approximately six times stronger than the SAGE sources, is described in Sec.~\ref{sec:CrSources}.

The procedure began with adding Ge carrier to each of the two zones and then installing the source into the center of the two-zone target volume, for an exposure of about 10~d. The source would then be moved to the calorimeter to measure its activity, while the Ga was pumped to the chemical reactors to perform the \nuc{71}{Ge} extraction. The extraction of the Ge (carrier and \nuc{71}{Ge}) was conducted over about a day. The GeH$_4$ gas was synthesized, mixed with Xe, and inserted into proportional counters. The gas was then counted for 60-150~d. The following subsections discuss these key experimental activities in further detail.
	
\subsection{The \nuc{51}{Cr} Source}\label{sec:CrSources}
	
\subsubsection{Source Fabrication}
Chromium has four isotopes. \nuc{50}{Cr} has a low natural abundance (4.35\%) and \nuc{53}{Cr} has a high neutron capture cross section. As a result, one must enrich the Cr in isotope 50, deplete in 53, to reach the desired activities~\cite{Cribier1988}. The Cr isotope enrichment for GALLEX and SAGE was done at the Kurchatov Institute by gas centrifugation of CrO$_2$F$_2$~\cite{TIKHOMIROV1992,POPOV1995}. The CrO$_2$F$_2$ was then hydrolyzed to Cr$_2$O$_3$ followed by reduction to metallic Cr. Impurities in Cr would activate while in the reactor, creating a potential health hazard and impacting the source strength measurement. Therefore, great care was taken to ensure no contamination occurred during processing.  To verify this, the samples were chemically analyzed by mass spectroscopy prior to irradiation.

SAGE extruded the Cr metal into rods, which were irradiated at the BN-350 fast breeder nuclear reactor in Aktau, Kazakhstan to produce a 516.6~kCi source~\cite{sage1999source}. GALLEX irradiated its Cr at Silo\`{e} at Grenoble as chips within a zircalloy tube~\cite{Cribier1996}. Both of these reactor facilities produce intense thermal neutron fluxes and allow for the loading of large samples for irradiation.

\begin{figure}
  \centering
  \includegraphics[width=0.5\columnwidth]{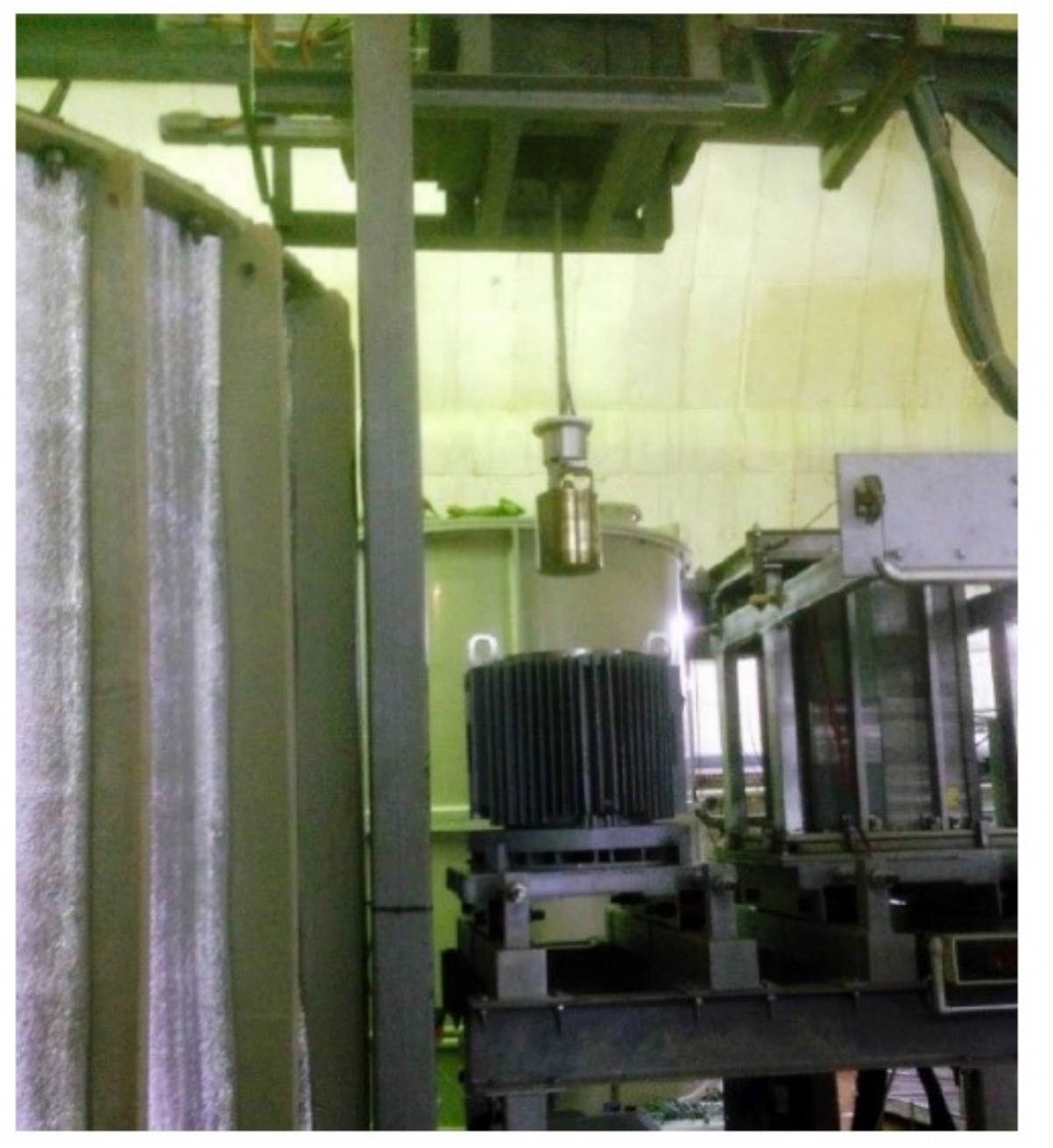}
  \caption{A photograph of the BEST source as it is being removed from its transport container. To the right side of the photo, the calorimeter can be seen.}
  \label{fig:BESTTSource}
\end{figure}

BEST used 4~kg of 97\%-enriched \nuc{50}{Cr} formed into 26 metal disks, which were irradiated for $\sim$100~d in the SM-3 reactor at the State Scientific Center Research Institute of Atomic Reactors, Dimitrovgrad, Russia. After irradiation the $3.1414\pm0.008$~MCi source was delivered to Baksan on 5 July 2019, with exposures beginning at 14:02 that same day. This was taken as the source strength reference time. The source is shown in Fig.~\ref{fig:BESTTSource}.

\subsubsection{Source Activity}
	 
\begin{table}[htp]
\caption{Summary of the source activities  and measured-to-predicted ratios for each of the six experiments. The experiments used different units to quote activities, therefore we give both here.
}
\begin{center}
\begin{tabular}{lccl}
\hline
Measurement 				&	Activity ($10^{15}$ Bq)		& Activity (MCi)									&Measured/Expected		\\
\hline\hline
SAGE Cr				 	&	\phantom0$19.11\pm0.22$	& $0.5166\pm0.0060$~\cite{sage1999source}			&$0.95\pm0.12$~\cite{sage1999source,sage2006argon}		\\
SAGE Ar				 	&	\phantom0$15.1\pm0.7$		& 	$0.409\pm0.002$~\cite{sage2006argon}			&	0.79$^{+0.09}_{-0.10}$~\cite{sage2006argon}\\
GALLEX Cr-1				 &	\phantom063.4$^{+1.1}_{-1.6}$~\cite{gallex1998source}		&	1.714$^{+0.03}_{-0.043}$			& 0.953$\pm0.11$~\cite{gallex2010reanalysis}\\
GALLEX Cr-2				 &	\phantom069.1$^{+3.3}_{-2.1}$~\cite{gallex1998source}		&	1.868$^{+0.09}_{-0.057}$			& 0.812$^{+0.10}_{-0.11}$~\cite{gallex2010reanalysis}	\\
BEST-inner				 	&	$116.23\pm0.03$			&	$3.1414\pm0.008$~\cite{BESTprl}			& $0.79\pm0.05$~\cite{BESTprl}	\\
BEST-outer				 	&	$116.23\pm0.03$			&	$3.1414\pm0.008$~\cite{BESTprl}			& $0.77\pm0.05$~\cite{BESTprl}	\\
\hline\hline
\end{tabular}
\end{center}
\label{tab:Activities}
\end{table}%

All three experiments used calorimetry as their primary, and most precise, tool to estimate the activity of the sources. All three also used a variety of additional methods to cross-check the activity determination. 

The activity of the SAGE \nuc{51}{Cr} source was measured three ways~\cite{sage1999source}. A calorimeter was built to measure the heating power of the source~\cite{Belousov1991}; the 320 keV $\gamma$-rays produced
following EC to the first excited state in $^{51}$V ($\sim$10\% branch) were counted; and the activity was estimated from reactor physics.

GALLEX~\cite{gallex1998source} also made several  measurements of source activity.  Samples were collected of the irradiated Cr chips and the 320-keV $\gamma$ was counted. This was done independently by three  groups (Saclay, Karlsruhe and BNL) along with inductively coupled plasma-atomic emission spectroscopy. The activity of the total source was also measured by calorimetry, and finally by measuring the \nuc{51}{V} content
of the source (Karlsruhe and BNL) after all irradiations were complete (at which time most of the \nuc{51}{Cr} had decayed).

The BEST calorimetry results~\cite{kozlova2020measurement,Gavrin_2021} were complimented by measurements of the $\gamma$ radiation from the source between exposures.  Figures~\ref{fig:BESTLabPhotos} and \ref{fig:BESTTSource} show photos of calorimeter.   The precision of the calorimetry, a technique with a long and successful history, combined with multiple supporting measurements, as described above, gives one confidence the the source intensities were very well determined.
	
\subsection{BEST Experimental Operations}
BEST performed 10 extractions from each of the two volumes for a total of 20 measurements of the \nuel\ flux.

\subsubsection{The Extractions}
The procedure for Ge extraction from metal Ga is described in detail in Ref.~\cite{sage1999solar}, with improvements employed in solar neutrino and source measurements after 1998 described in Ref.~\cite{sage2006argon}.   The extraction efficiency is measured by introducing a Ge carrier isotope to the Ga target at the beginning of each neutrino exposure.  In the procedure followed since 2005,  2.4 $\mu$mol of Ge enriched in either  \nuc{72}Ge (92\%) or \nuc{76}Ge (95\%) is added.  The contents of the reactor are then stirred to thoroughly disperse the carrier throughout the target. At the end of the exposure, an extraction solution consisting of HCL and 30\% H$_2$O$_2$ is added and the Ga is intensively stirred. This causes the Ga to form into fine droplets which are covered with a Ga oxide film.  This film prevents fusion of the droplet and holds the Ga as a fine emulsion. The dissolved Ge in the Ga migrates to the surface of the droplets, oxidizes, and is incorporated into the oxide film. Once the H$_2$O$_2$ is consumed, the emulsion breaks down.  To dissolve the oxide containing the Ge, a quantity of 7 M HCl is added and the Ga briefly stirred.  This solution is decanted and concentrated by evaporation to a volume for sweeping. The Ge is swept from this volume as GeCl$_4$, which is volatile and thus can be swept out with a flow of air. The Ge is extracted into CCl$_4$ and then back extracted into low-tritium water. The process is repeated three times to concentrate the Ge into a small volume of water.  A much more detailed discussion of these chemical procedures can be found in Ref.~\cite{sage1999solar}. 

Based on tracer recovery, the overall Ge extraction efficiency for the BEST runs was 98\%.

\subsubsection{GeH$_4$ Synthesis}
GALLEX, SAGE, and BEST followed similar procedures for synthesizing the GeH$_4$ gas~\cite{Anselmann1992,sage1999solar}. The Ge-loaded water from the extraction has a final volume of about 100~ml. NaOH is added to adjust the pH and the solution is placed in a reduction flask. Low-tritium NaBH$_4$ dissolved in low-tritium water is added and the mixture is heated to 70~C. At this temperature the Ge is reduced by the NaBH$_4$, making GeH$_4$. The produced H$_2$ and flowing He sweep the GeH$_4$ into a chromatography unit where it is captured in a cold trap. After the reaction completes, the column temperature is raised and the GeH$_4$ is eluted with the He and frozen on another trap.  It is then released, mixed with Xe and added into a miniture proportional counter.

The overall synthesis efficiency for the BEST runs was 96\%.
	
\subsubsection{The Proportional Counters}
The small ($\sim$0.5~cm$^3$) proportional counters used by BEST were identical to those SAGE employed after 2001 and also similar to those of GALLEX. The counters had a thin layer of a carbon deposited on the inner surface of a quartz body using thermal decomposition of isobutane. This layer serves as the cathode and minimized dead volume. The special design, with the walls rounded inwards near the cathode ends, minimizes edge effects. Connections to the cathode and anode were made of molybdenum band, which provided a good gas seal and guaranteed stability of amplification. This design had lower background and higher volume efficiency than the earlier SAGE design, while maintaining stable high gas amplification and good energy resolution. The counters were manufactured from radiopure materials. The counter bodies were fabricated of synthetic quartz (Suprasil\textregistered ~\cite{Heraeus}. The thickness of this body wall was etched with hydrofluoric acid to about 200~$\mu$m. This kept the background from the Suprasil\textregistered very low.  

Earlier SAGE counters used a zone-refined iron sleeve as the cathode. These were low background, but had a dead volume behind the sleeve. The GALLEX counters were similar to the early SAGE counters with an iron or silicon cathode sleeve and tungsten anode wire with a body made of Suprasil\textregistered~\cite{Anselmann1992}.

\begin{figure}
  \centering
  \includegraphics[width=0.15\columnwidth,angle=90]{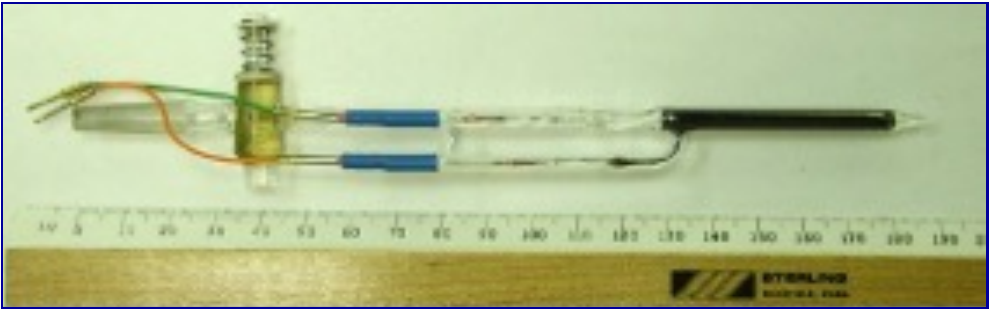}
  \caption{A photograph of a BEST proportional counter.}
  \label{fig:CounterPhoto}
\end{figure}

\subsubsection{Waveform Analysis and Likelihood Fits}	
\begin{figure}
  \centering
  \includegraphics[width=0.5\columnwidth]{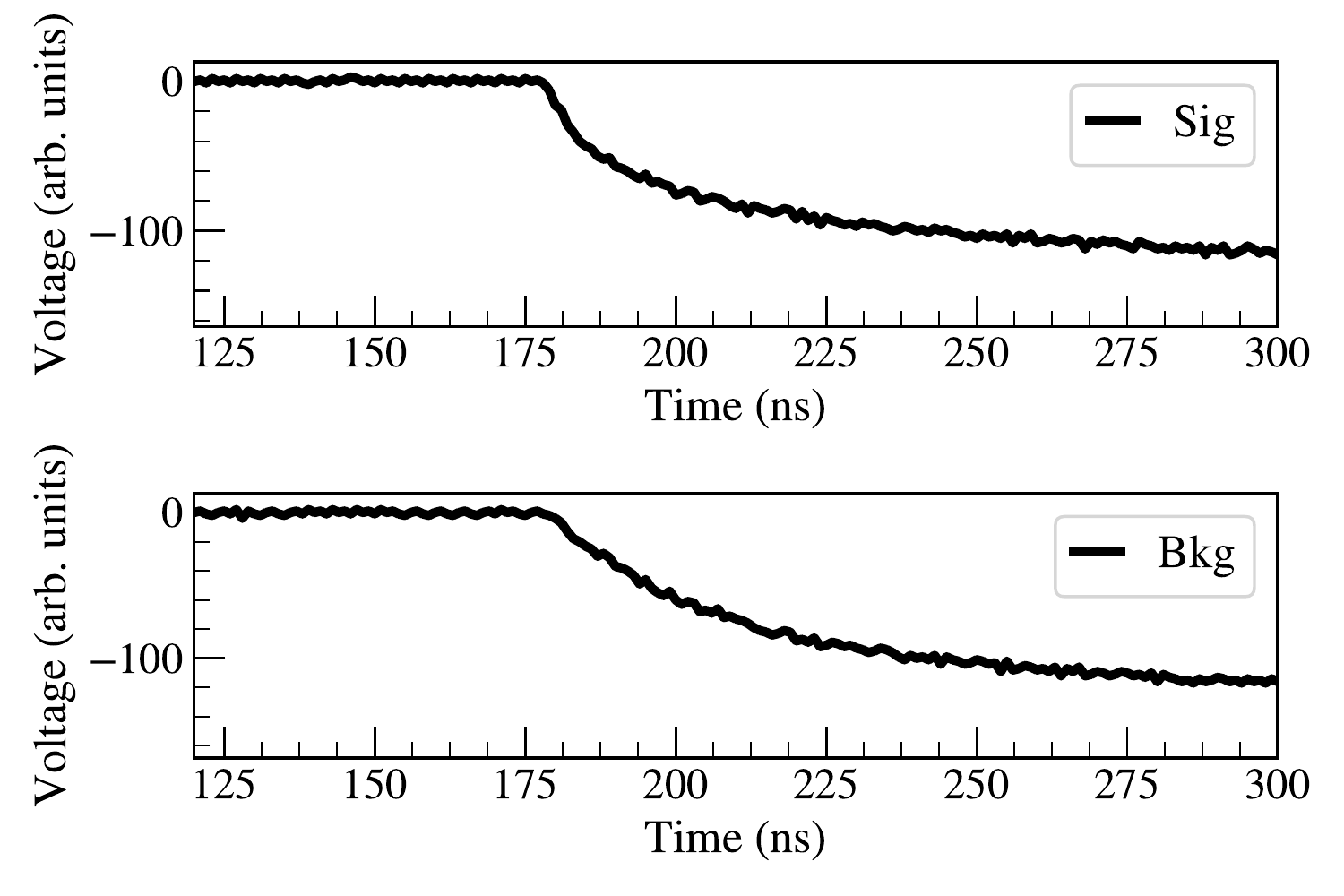}
  \caption{Top: A candidate signal pulse. Bottom: A candidate background pulse.}
  \label{fig:CounterPulses}
\end{figure}

\begin{figure}
  \centering
  \includegraphics[width=0.5\columnwidth]{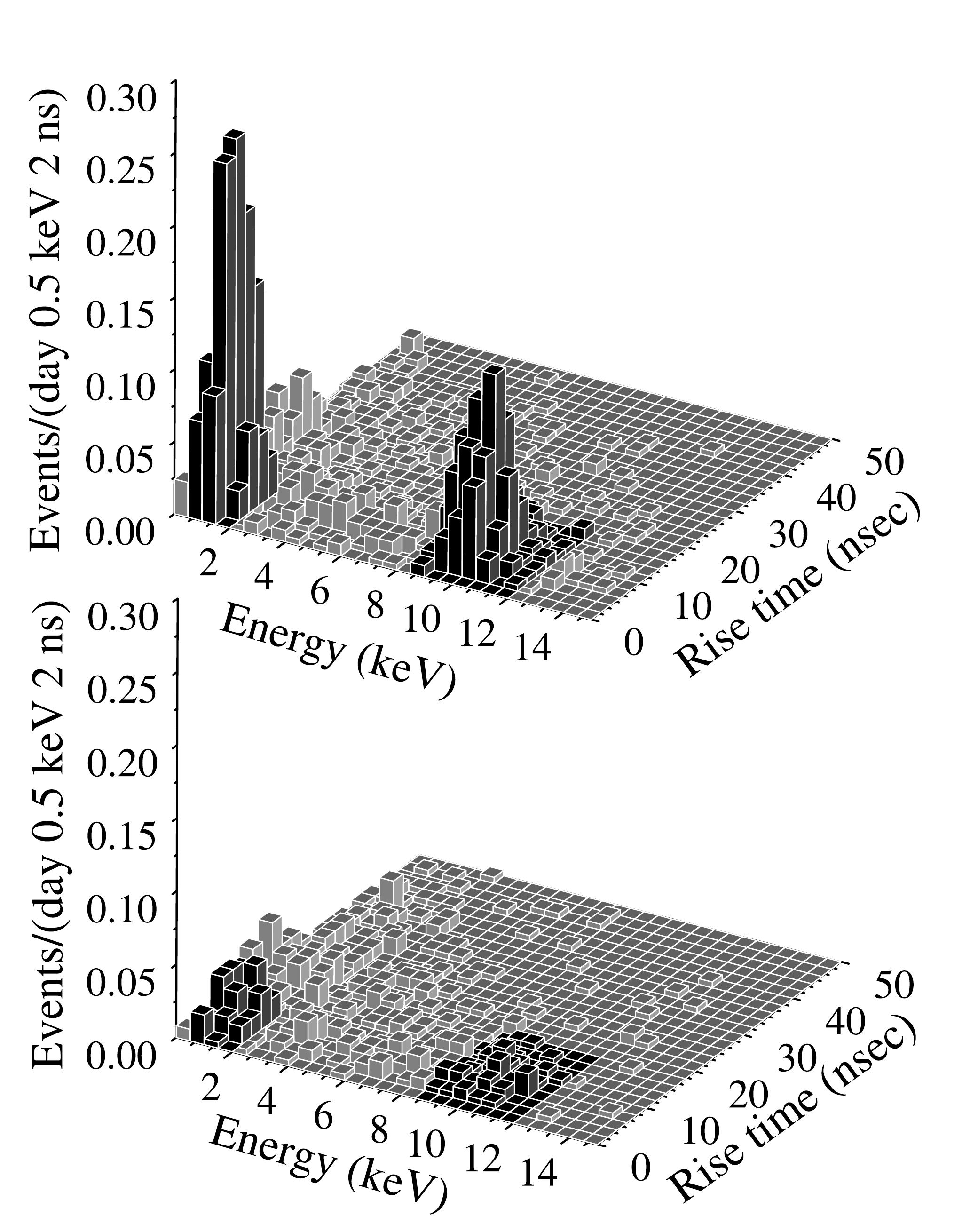}
  \caption{The candidate events from the BEST experiment separated into time bins after start of counting~\cite{BESTprc}. Top: Energy vs rise-time histogram of all events of the outer target after the shield-open cut observed in all ten exposures during the first 30 days after extraction. The live time is 249~days, and 1387~events are shown. Bottom: The same histogram for the 504~events that occurred during an equal live-time interval beginning at 40~days after extraction. }
  \label{fig:LEGOPlot}
\end{figure}

A  great deal of information is encoded in the rise times of the digitized waveforms from the proportional counters. The Auger electrons arising from the decay of \nuc{71}{Ge} deposit all their energy in a small volume. As the charge from this point-like energy deposition arrives at the central counter wire, it produces a pulse with fast rise time, as shown in the top panel of Fig.~\ref{fig:CounterPulses}. In contrast, background events from $\beta$ particles or Compton electrons may deposit a similar amount of energy, but produce an extended track, so that the ionization arrives at the central wire over an interval. This generates a pulse with a slower rise time, as shown in the bottom panel of Fig.~\ref{fig:CounterPulses}. This difference has been exploited to improve the signal-to-background ratio~\cite{ALTMANN1996,elliott1990analytical}. The power of rise-time techniques to distinguish signal from background is apparent from Fig.~\ref{fig:LEGOPlot}. The two panels show data taken early during counting when \nuc{71}{Ge} has not fully decayed away and hence shows its signature, and late in the counting after the \nuc{71}{Ge} has decayed and only background remains.

Events selected by energy and rise time are used in a likelihood fit~\cite{Cleveland1983}. Here the time of each event is used to determine its probability of being a signal event or background. The number of signal events that maximizes the likelihood is used to determine the production rate. To determine the quality of fit, Cramer-von Mises and Anderson-Darling statistical tests were used~\cite{Cleveland1998GoodnessOfFit}.

The final BEST results are displayed in Fig.~\ref{fig:ProdRatio} along with those obtained in the four earlier solar neutrino calibration experiments. The significantly improved precision of the BEST measurements is clear. The measured-to-expected ratios found for the BEST inner and outer vessels are $R_{in} = 0.79\pm0.05$ and $R_{out} = 0.77\pm0.05$, respectively. These values differ significantly from unity, but agree with each other within uncertainties, revealing no tell-tale sign of oscillations. The results for all six measurements are given in Table~\ref{tab:Activities}. The auxiliary tests that have been performed to check whether an experimental artifact is the cause of the observed deficits are discussed in Sec.~\ref{sec:Auxiliary}.  Cross section uncertainties are discussed in Sec.~\ref{sec:CrossSection}. The various nuclear and atomic physics inputs that might affect the result are discussed in Sec.~\ref{sec:NPInput}.

Some of the interest in BEST and other Ga calibration tests comes from suggestions that \oscil\ might be responsible for the low values of $R$, as well as other neutrino anomalies. This possibility and the tension with other experiments is discussed in Sec.~\ref{sec:Oscillations}.

\section{Auxiliary Experimental Tests}\label{sec:Auxiliary}
The SAGE and GALLEX programs considered various steps in the Ge recovery and counting for which the efficiencies might have been overestimated. No evidence of such was found. The following subsections describe the tests that were performed.
	
\subsection{Extraction and Synthesis Efficiency}
The GALLEX extraction efficiency is based on the volatility of GeCl$_4$ with Ge in the tetravalent state. Efficient extraction of Ge(IV) carrier and \nuc{71}{Ge}(IV) was confirmed by a number of tests. One such test considered whether \nuc{71}{Ge} produced by \nuc{51}{Cr} \nuel\ capture retained the molecular form for extraction. GALLEX performed a \textit{hot-atom chemistry} test with \nuc{71}{As}~\cite{gallex1998arsenic}. Hot-atom chemistry refers to the feature that the produced \nuc{71}{Ge} has a recoil energy resulting from \nuel\ capture and subsequent $\beta$ emission. As the recoil energy is comparable to the $\sim$3-4~eV chemical binding energy, the Ge-Cl bond could be broken, resulting in a depressed extraction efficiency. Also, EC produces an inner shell vacancy that, as it fills, can produce shake-off electrons altering the charge state, again resulting in molecule breakup.    If the Ge ends up in a non-extractable form such as Ge(II) instead of the expected Ge(IV), the carrier recovery measurements might not be an accurate measure of the extraction efficiency.

One can check hot-chemistry effects using \nuc{71}{As}, which decays by EC and $\beta^+$, producing  \nuc{71}{Ge} with  kinematics resembling those resulting from  \nuc{71}Ga$(\nu_e,e^-)$\nuc{71}{Ge} for \nuc{51}{Cr} source $\nu_e$s.  By adding a known amount of \nuc{71}{As} ($\tau_{1/2}$=2.72~d) and counting the number of extracted \nuc{71}{Ge} atoms produced, GALLEX performed a high-statistics
check of potential hot-chemistry effects.  The recovery of \nuc{71}{Ge} was 100$\pm$1\%, indicating that the produced \nuc{71}{Ge} ends up as volatile extractable GeCl$_4$. As there is no known technique for dissolving and stirring As within Ga metal, similar tests for SAGE and BEST have not been performed. 

Instead, SAGE performed Ga metal hot-atom chemistry tests by doping the detector with radioactive \nuc{70,72}{Ga} produced by neutron activation~\cite{sage1999source}. 
This checked whether introduction of the carriers \nuc{70,72}{Ge} via {\it in situ} decay, and thus with significant recoil energy, would influence the efficiency of their recovery. No change in recovery efficiency was seen.
However, as the maximum recoil energy of \nuc{70}{Ge} after $\beta$ decay, 32~eV, is larger than both the 20 eV recoil of \nuc{71}{Ge} after \nuc{51}{Cr} \nuel\ capture and the 6.1 eV recoil after pp \nuel\ capture, there is not
a precise equivalence between the carrier kinematics and those produced in the neutrino capture reactions.

On the Earth's surface, \nuc{68}{Ge} ($\tau_{1/2} \sim$271 d)  is produced cosmogenically within Ga. When the Ga was initially brought underground, extractions were conducted by SAGE to remove \nuc{68}{Ge}. The reduction of this isotope during these extractions followed that of the Ge carrier~\cite{sage1999source}. GALLEX also did extractions to remove \nuc{68}{Ge}.  A small fraction of the \nuc{68}{Ge} was retained: they attributed the extraneous activity to some trace impurity
intracting with the \nuc{68}Ge, releasing it slowly~\cite{Anselmann1992}. They note the effect is very small and only observable because of the very high initial level of \nuc{68}{Ge}.

SAGE prepared a sample of carrier doped with a known number of \nuc{71}{Ge} atoms, adding the isotope to a reactor containing seven tons of Ga. Three extractions were performed and the measured
rate was as expected from the stable carrier determination~\cite{sage1999source}.
	
The pp solar neutrino flux measurements of SAGE and GALLEX/GNO agree, but also can be compared to the result from Borexino. If the radiochemical Ga experiments had an efficiency lower than claimed, this would be revealed by
a higher rate in Borexino's event-by-event detection. The Ga result, $(6.0\pm0.8)\times10^{10}$/cm$^2$s~\cite{abdurashitov2009measurement}, agrees with that from Borexino, $(6.1\pm0.5^{+0.3}_{-0.5})\times10^{10}$/cm$^2$s~\cite{agostini2018comprehensive}.  However, the comparison is not definitive due to the $\sim$12\% uncertainty of each measurement.	

\subsection{Counter Efficiency}
A variety of tests were performed by SAGE, GALLEX and BEST to ensure their counter efficiencies were estimated correctly. SAGE performed a counter efficiency test with \nuc{69}{Ge}~\cite{Abdurashitov1995PandC}. This isotope decays 27\% of the time by EC with a signal identical to that of \nuc{71}{Ge}. The remainder of the decays are EC followed by emission of a 1106-keV $\gamma$. By placing a proportional counter near a HPGe detector, both the $\gamma$s and Auger emissions could be measured, confirming the Auger detection efficiency. SAGE also measured the volume efficiency using counters filled with \nuc{37}{Ar} and \nuc{71}{Ge}. The radioactive gas mixtures were measured in a typical counter then transferred to a  large counter specifically designed for very high efficiency \cite{sage1999solar}. These radioactive-gas fills also provided data to verify the peak counting efficiency. The volume efficiency was corrected for gas pressure and mixture (GeH$_4$ fraction). GALLEX used \nuc{71}{Ge}-filled counters \cite{Hampel1985ICRC,Plaga1991} to establish their efficiencies and optimize counter design. BEST measured the volume efficiency of each counter with \nuc{37}{Ar} and \nuc{71}{Ge} after the experimental measurements were completed. These measurements also determined the peak selection efficiency and the rise-time cut efficiency~\cite{BESTprc}.

\subsection{Detector Effective Path Length and Distribution}
This effective neutrino path length $\langle L \rangle$ through each zone of the BEST detector can be computed from Eq. (\ref{eq:third}).  
$\langle L \rangle$, target
Ga mass, and the distribution $L(D)$ were evaluated by Monte Carlo integration, for both the near and far zones.  The as-built geometry for BEST was used to define the integration volumes.
Uncertainties in these quantities arise from the precision of the as-built measurements, the statistical precision of the Monte Carlo integration, and the comparison between the calculated and measured mass~\cite{BESTprc}. The total uncertainty is estimated to be $\pm$0.3\%, and thus the geometry is a negligible uncertainty in computing any of these quantities.  The values for $\langle L \rangle$ are given in Table~\ref{tab:MeanFreePath}. The two volumes were designed to provide similar values for $\langle L \rangle$ and thus similar event rates, in the absence of oscillations.

\begin{figure}
  \centering
  
\includegraphics[width=0.5\columnwidth]{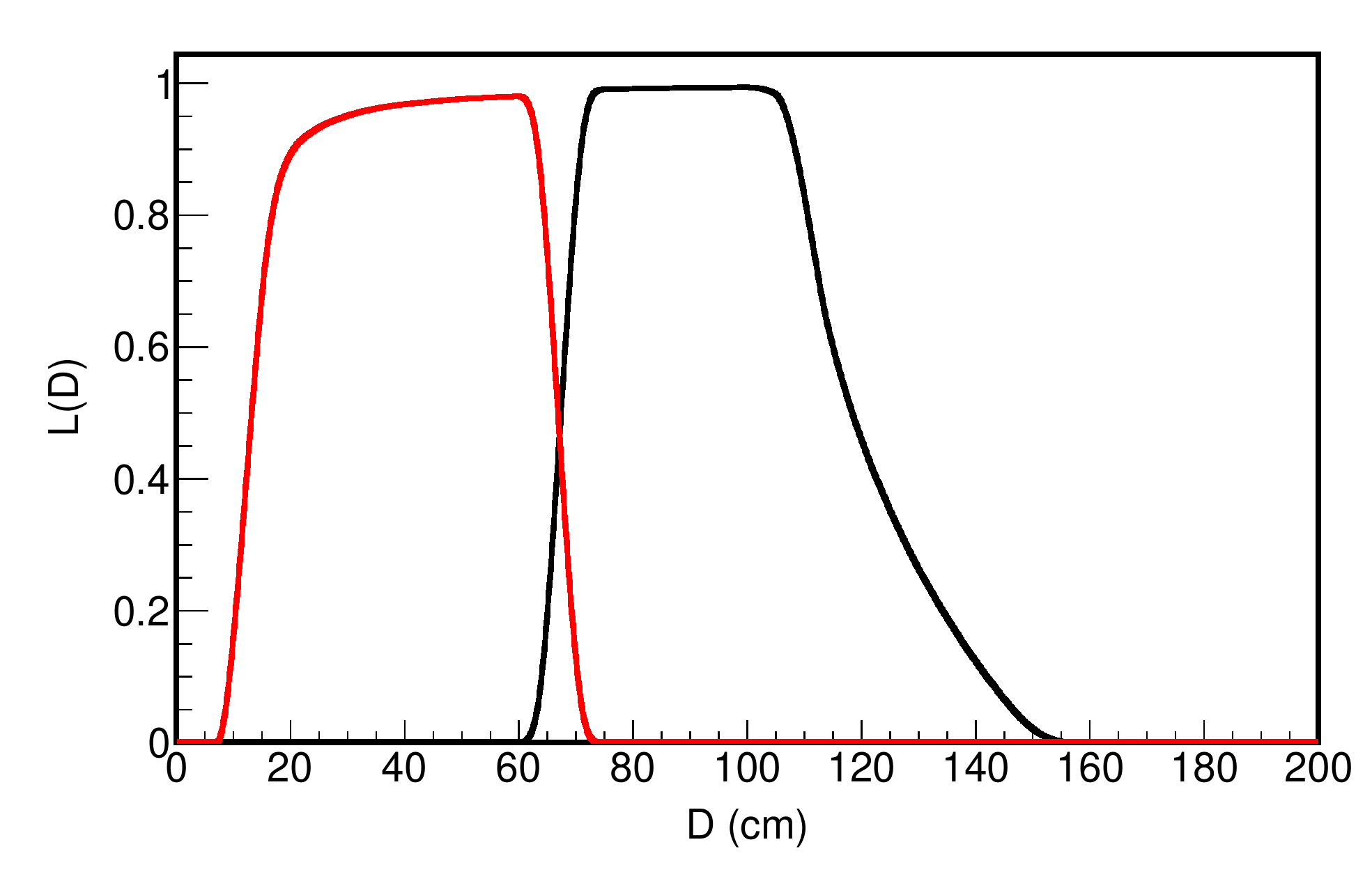}
\caption{The dimensionless distribution $L(D)$ which encodes detector and source geometry needed to compute the distribution of neutrino captures within the Ga volumes as a 
function of baseline (the distance $D$ between the points where the neutrino is produced in the source and detected in the Ga volume).  The total rate in the presence of oscillations is proportional to the
integral over the convolution  of $L(D)$ with the oscillation probability $P_{ee}(E_\nu^i,D)$ (see Sec. \ref{sec:D}).  Figure courtesy of Ralph Massarczyk.}
  \label{fig:OscilLength}
\end{figure}

In Fig. \ref{fig:OscilLength} we plot the dimensionless quantity $L(D)$, which encodes the source and detector geometry effects needed in determining event rates in the detector as a
function of baseline, the distance between the points of neutrino production in the \nuc{51}Cr source and detection in the Ga.  See the discussion in Sec. \ref{sec:D}.  In the
presence of oscillations, the event rate $r$ is proportional to the integral over possible baselines $D$ of the convolution of $L(D)$ with the oscillation probability $P_{ee}(E_\nu^i,D)$.
~\\

\begin{table}[htp]
\caption{The mean free paths for the various measurements and the range of values for the oscillation length. The oscillation length ranges are our estimates based on the given geometry.}
\begin{center}
\begin{tabular}{lcc}
\hline
Measurement 				&	$\langle L \rangle$ (cm)			& Oscillation Length Range (cm)									\\
\hline\hline
SAGE Cr				 	&	$72.6\pm0.2$~\cite{sage1999source}			& 	0-109		\\
SAGE Ar				 	&	$72.6\pm0.2$~\cite{sage1999source}			& 	0-109			\\
GALLEX Cr-1				 &	$190\pm1$~\cite{gallex1995firstCr}			&	0-194		\\
GALLEX Cr-2				 &	$190\pm1$~\cite{gallex1995firstCr}			&	0-194			\\
BEST-inner				 &	$52.03\pm0.18$~\cite{BESTprc}		&	0-67				\\
BEST-outer				 &	$54.41\pm0.18$~\cite{BESTprc}		&	67 -152 				\\
\hline\hline
\end{tabular}
\end{center}
\label{tab:MeanFreePath}
\end{table}%

\section{The \nuc{71}{Ga} Capture Cross Section}\label{sec:CrossSection}
\begin{figure}
  \centering
  \includegraphics[width=0.5\columnwidth]{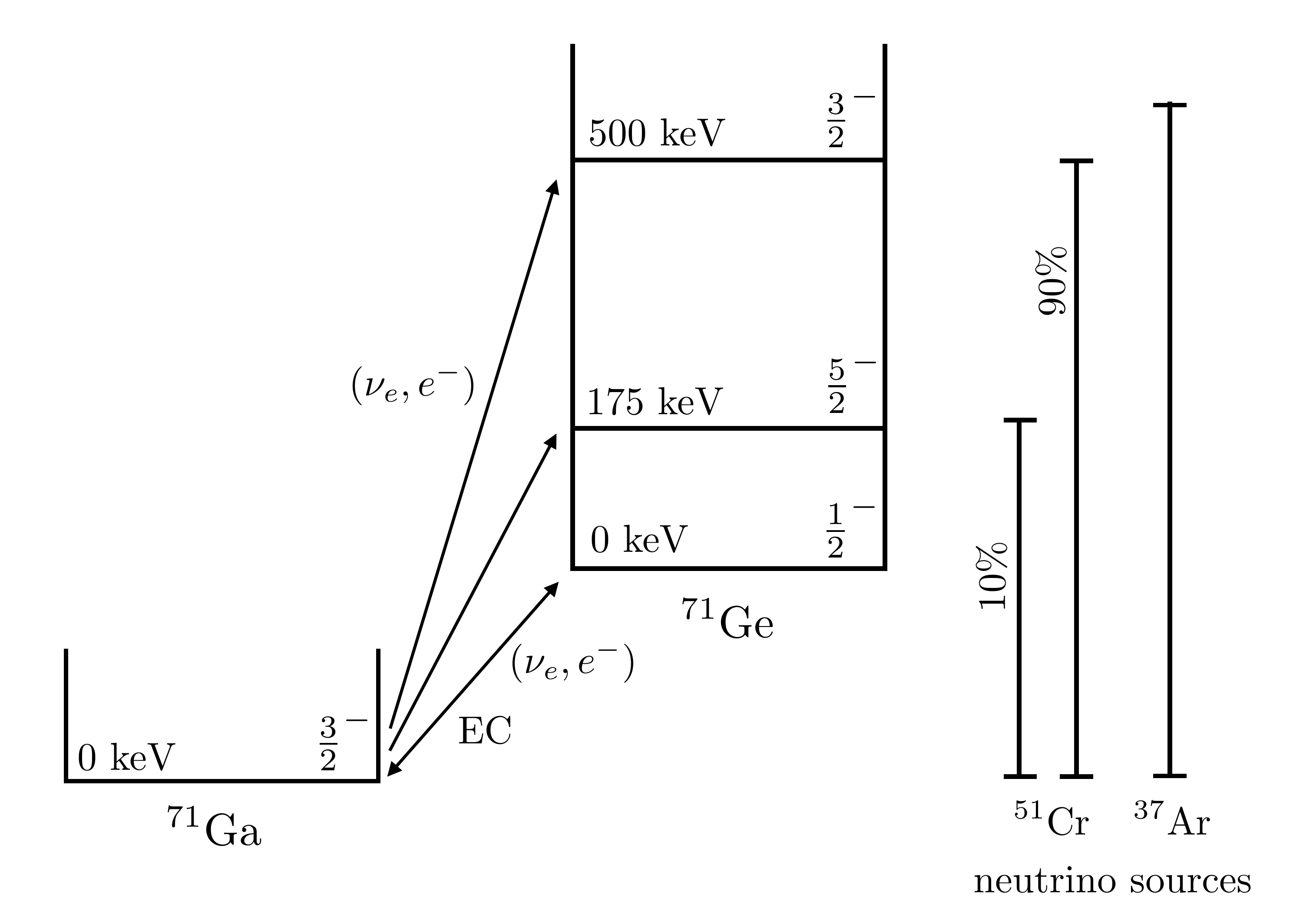}
  \caption{Level diagram for \nuc{71}Ga$(\nu_e,e^-)$\nuc{71}Ge showing the states contributing to the absorption of \nuc{51}Cr and \nuc{37}Ar EC neutrinos.}
  \label{fig:Nuclear}
\end{figure}

A critical issue in the analysis of the BEST and earlier Ga neutrino source calibrations is the cross section for \nuc{71}Ga$(\nu_e,e^-)$\nuc{71}Ge,
a common systematic in all of the experiments.  The nuclear physics is shown in Fig. \ref{fig:Nuclear}. 
Here we summarize work very recently completed in which the four aspects of this problem
were re-examined \cite{cross_2023}
\begin{enumerate}
\item The strength of the transition from the $\textstyle{3 \over 2}^-$ ground state of \nuc{71}Ga to the $\textstyle{1 \over 2}^-$ ground state of \nuc{71}Ge.  This transition dominates the neutrino absorption and is tightly constrained by
the known electron-capture lifetime of \nuc{71}Ge.
\item The associated partial neutrino cross section \nuc{71}Ga(gs)$(\nu_e,e^-)$\nuc{71}Ge(gs), computed from EC rate after a series of electroweak
corrections are made.
\item Cross section contributions of two kinematically accessible excited states in \nuc{71}Ge, at 175 keV ($\textstyle{5 \over 2}^-$)
and 500 keV ($\textstyle{3 \over 2}^-$).
\item The line neutrino spectra from the EC sources  $^{51}$Cr and $^{37}$Ar that are folded with the cross section, in rate estimations.
\end{enumerate}
The fourth point was addressed early in this overview:  the cross sections presented below represent weighted averages over the neutrino lines 
from each source, with the weights determined by the measured EC branching ratios.  The needed data are given in Table \ref{tab:SourceIsotopes}.

\subsection{The Ground State Transition Strength}
 The allowed strength for the neutrino-driven transition from \nuc{71}Ga to the ground state of \nuc{71}Ge is
\begin{equation}
\mathrm{B^{(\nu,e)}_{GT}}(\mathrm{gs}) \equiv  {1 \over 2j_i +1} \left| \langle j_f^\pi={\textstyle {1 \over 2}}^- || \sum_{1=1}^A \sigma(i) \tau_-(i) ||j_ i^\pi={\textstyle {3 \over 2}}^- \rangle \right|^2  
\end{equation}
where $\sigma(i)$ is the Pauli spin matrix, $\tau_-(i)$ is the isospin lowering operator, and $||$ denotes a matrix element reduced in angular momentum.  As shown in \cite{cross_2023}, the Gamow-Teller (GT) transition strength $\mathrm{B_{GT}}$ can be extracted from the precisely measured electron-capture half-life of \nuc{71}Ge \cite{hampel1985}
\[  \tau_{1 \over 2} [^{71}\mathrm{Ge}] = 11.43 \pm 0.03~\mathrm{d} \]
through the relation
\begin{equation}
 \omega = {\mathrm{ln}[2] \over \tau_{1 \over 2} } = {G_F^2 \cos^2{\theta_C} \over 2 \pi} ~|\phi^{av}_{1s}|^2 ~ E_{\nu,1s}^2~ { \textstyle \left[2(1+\epsilon_o^{1s}) (1+ {P_L+P_M \over P_K} )\right]}~ ~ g_A^2~ [2 ~\mathrm{B^{(\nu,e)}_{GT}}(\mathrm{gs})]~[1+g_{v,b}]_{EC} ~[ 1 + \epsilon_\mathrm{q}]  
 \label{eq:sigmao}
 \end{equation}
 This expression gives the total EC rate in terms of the partial rate for $1s$-capture.  The terms in Eq. (\ref{eq:sigmao}) are
 \begin{enumerate}
 \item  The weak couplings.   This expression was evaluated in \cite{cross_2023} using Particle Data Group values for the Fermi coupling constant $G_F$ and the Cabibbo angle $\theta_C$ and the Perkeo II value for the axial  vector coupling $g_A$.
 \item The energy of the emitted neutrino $E_{\nu,1s}$, computed from the EC Q-value and $1s$ binding energy of 10.37 keV
  \[ Q_{EC}=M[^{71}\mathrm{Ge}]-M[^{71}\mathrm{Ga}] = 232.443 \pm 0.093~ \mathrm{keV} \]
\[ Q_{EC}= E_{\nu,1s} + 10.37 \mathrm{~keV} ~\Rightarrow E_{\nu,1s}  =222.1 \pm 0.1 \mathrm{~keV}. \]
\item The wave function probability, averaged over the nuclear volume, for a $1s$ electron, $|\phi^{av}_{1s}|^2$.  This quantity is taken from atomic many-body theory, and its uncertainty is addressed in the next section.
\item The factor in the first square bracket relates single-electron $1s$ capture rate to the total EC rate.  It includes a) a factor of two, as there are two $1s$ electrons;
b) an exchange and overlap correction, taken from theory, that accounts for the fact that an instantaneous hole in the atomic cloud of \nuc{71}Ge does not overlap
precisely with a similar hole in the daughter nucleus \nuc{71}Ga and c) the contributions from L and M EC capture relative to K capture, expressed in terms of the
experimentally known capture probabilities $P_K$, $P_L$, and $P_M$.
\item The second square bracket expresses the allowed matrix element for EC in terms of that for $(\nu_e,e^-)$, B$_\mathrm{GT}^\mathrm{EC}= 2 ~\mathrm{B^{(\nu,e)}_{GT}}(\mathrm{gs})$.
\item The factor $[1+g_{v,b}]_{EC}$ is the contribution from radiative corrections (the exchange of virtual photons and bremsstrahlung).
\item The factor $[ 1 + \epsilon_\mathrm{q}]$ represents contributions beyond the allowed approximation.  This contribution is dominated
by the term linear in the three-momentum transfer $q$ arising from interference between the allowed amplitude and weak magnetism.  While the weak magnetism contribution is constrained by
the known isovector magnetic moment, there are additional contributions that must be taken from nuclear theory \cite{cross_2023}.
 \end{enumerate}
 
 The various terms in Eq. (\ref{eq:sigmao}) have uncertainties, which are discussed in detail in \cite{cross_2023} and, partially, in the next section of this paper.
 One finds that the gs $\leftrightarrow$ gs transition strength for $(\nu_,e^-)$ is constrained by the known EC rate to about 1\%, when all such uncertainties
 are considered
 \[ \mathrm{\tilde{B}^{(\nu,e)}_{GT}}(\mathrm{gs}) \equiv  \mathrm{B^{(\nu,e)}_{GT}}(\mathrm{gs})~[1+g_{v,b}]_{EC} = 0.0864 \,  \pm \,  0.0010~~ (95\% \, \mathrm{C.L.}) \]
 
 \subsection{The Ground State Neutrino Capture Cross Section}
The cross section for  \nuc{71}Ga$(\nu_e,e^-)$\nuc{71}Ge(gs) can be expressed in terms of the allowed matrix element defined above,
\begin{equation}
\sigma_\mathrm{gs} = \frac{G^2_F \cos^2{\theta_C}}{\pi}\,  p_e E_e \mathcal{F}(Z_f, E_e) \,  g^2_A  \, \tilde{\mathrm{B}}_\mathrm{GT}^{(\nu,e)}(\mathrm{gs})  \, \frac{[1+g_{v,b}]_{(\nu,e)}}{[1+g_{v,b}]_{EC}} \, [1+\epsilon_q].
\label{eq:GSCrossSection}
\end{equation}
The  terms in Eq. (\ref{eq:GSCrossSection}) are
\begin{enumerate}
\item $E_e$ and $p_e$ are the energy and three-momentum of the produced electron, related to the Q-value by
\[ E_e = E_\nu-Q_{EC} +m_e -0.09 \, \mathrm{keV} \]
where the last term on the right is a small correction for the energy lost to electronic rearrangement, following the neutrino reaction.
\item $\mathcal{F}(Z_f, E_e)$ is the correction for the effects of Coulomb distortion of the outgoing electron. It corresponds to the Dirac solution for a nuclear
charge distribution, here described as a Fermi distribution constrained to reproduce the experimental r.m.s. radius, with additional corrections to account for 
atomic screening.
\item The ratio $[1+g_{v,b}]_{(\nu,e)}/[1+g_{v,b}]_{EC}$ accounts for the differential effects of radiative correction on the inverse reactions of EC and
$(\nu_e,e^-)$.  This difference, generated by the bremsstrahlung contribution, was calculated as in \cite{Kurylov}.
\item The factor $[1+\epsilon_q]$ is the correction for forbidden contributions, again dominated by the interference term involving weak magnetism. 
\end{enumerate}

When these terms are combined and associated errors propagated, one finds for the \nuc{51}Cr and \nuc{37}Ar sources \cite{cross_2023}
\begin{equation}
\sigma_{gs} = \left\{ \begin{array}{ll} (5.39 \pm 0.06) \times 10^{-45} ~\mathrm{cm}^2 & ^{51}\mathrm{Cr} \\  (6.45 \pm 0.07) \times 10^{-45} ~\mathrm{cm}^2 & ^{37}\mathrm{Ar} \end{array} \right. ~~(95\% \, \mathrm{C.L.})
\end{equation}

 \subsection{Excited-State Contributions to the Neutrino Capture Cross Section}
 The strength of the well-determined ground state cross section is already sufficient to generate a Ga anomaly.  The contributions from the 
 $\textstyle{5 \over 2}^`$ (175 keV) and $\textstyle{3 \over 2}^-$ (500 keV) excited states exacerbate the anomaly, but unlike the ground state transition,
 are not constrained by known weak rates.  As described in \cite{cross_2023}, early efforts to estimate the excited-state contributions utilized nuclear
 systematics and shell model (SM) calculations, producing uneven results and thwarting efforts to assign uncertanties.  With the establishment the 
 forward-angle (p,n) scattering as a reliable probe of the B$_\mathrm{GT}$ strength distribution, this became the method of choice to constrain the
 two excited states.  The (p,n) \nuc{71}Ga strength distribution was measured by Krofcheck et al. \cite{Krofcheck},
 \begin{eqnarray}
\label{eq:BGT}
\mathrm{B_{GT}^{(p,n)}} \left[^{71}\mathrm{Ga} (\mathrm{gs}) \rightarrow  {}^{71}\mathrm{Ge} (\textstyle{5 \over 2}^-;175 \mathrm{~keV}) \right] \lesssim 0.005~~~~~~~~~~ \nonumber \\
\mathrm{B_{GT}^{(p,n)}} \left[^{71}\mathrm{Ga}(\mathrm{gs}) \rightarrow  {}^{71}\mathrm{Ge}(\textstyle{3 \over 2}^-;500 \mathrm{~keV})  \right] = 0.011 \pm 0.002 
\label{eq:Krofcheck}
\end{eqnarray}
These results were then used by Bahcall \cite{Bahcall1997} and others to estimate the cross sections to the two excited states.

Earlier Hata and Haxton \cite{Hata1995} had pointed out that the (p,n) reaction, while successful when used to map the broad profile of B$_\mathrm{GT}$
strength, is not a reliable probe of individual weak GT transitions, such as those in Eq. (\ref{eq:Krofcheck}), unless additional corrections are made.   By comparing transitions
with known weak strengths with the corresponding forward-angle (p,n) cross sections, an effective operator was empirically determined \cite{Watson}.
It includes, in addition to the GT operator, a subdominant contribution from a tensor operator $\hat{O}^{J=1}_\mathrm{T}$, 
\be
M^\mathrm{(p,n)} \equiv M_\mathrm{GT} + \delta M_\mathrm{T} ~~~~~~~ M_\mathrm{T} \equiv \langle J_f \alpha_f ||  \hat{O}^{J=1}_\mathrm{T}  || J_i \alpha_i \rangle~~~~~~~
\hat{O}^{J=1}_\mathrm{T}=\sqrt{8 \pi}  \sum_{j=1}^A  \left[Y_2(\Omega_j) \otimes \sigma(j) \right] _{J=1} \tau_+(j)~~~~~~
\label{eq:MT}
\ee
where $\delta \sim 0.1$, so that
\be
\mathrm{B_{GT}^{(p,n)}} = {1 \over 2 j_i +1} | \langle J_f \alpha_f || M^\mathrm{(p,n)} || J_i \alpha_i \rangle|^2~ .
\ee
It was stressed in \cite{Hata1995,Haxton1998} that \nuc{71}Ga$(\nu_e,e^-)$\nuc{71}Ge might be quite sensitive to the tensor contribution, 
as the \nuc{71}Ga $\textstyle{3 \over 2}^-$ ground state transition to the 175 keV $\textstyle{5 \over 2}^-$ excited state might be dominated by $\ell$-forbidden
$2p_{3/2} \leftrightarrow 1f_{5/2}$ amplitude.  Thus while $\delta \sim 0.1$, the tensor operator could easily dominate the transition.

Equation (\ref{eq:MT}) states the GT strength can still be measured through forward-angle (p,n) measurements, provided one subtracts the contribution from $M_T$.   While analyses were done in \cite{Hata1995,Haxton1998}, the recent work of \cite{cross_2023} was the first to adequately assess the whether $M^\mathrm{(p,n)}$ could quantitatively account
for known weak transition rates.  In this work, a) the transitions were carefully selected so that only data with well-established errors were included;
b) the estimates of $M_T$, which must be taken from theory, were computed for multiple interactions, to provide a measure of that uncertainty; and c) the pattern of the
results were displayed as a function of the B$_\mathrm{GT}$ strength, to clearly display the effects of the subdominant amplitude.  The many details of this
analysis are given in \cite{cross_2023}.

\begin{figure}
  \centering
  \includegraphics[width=0.75\columnwidth]{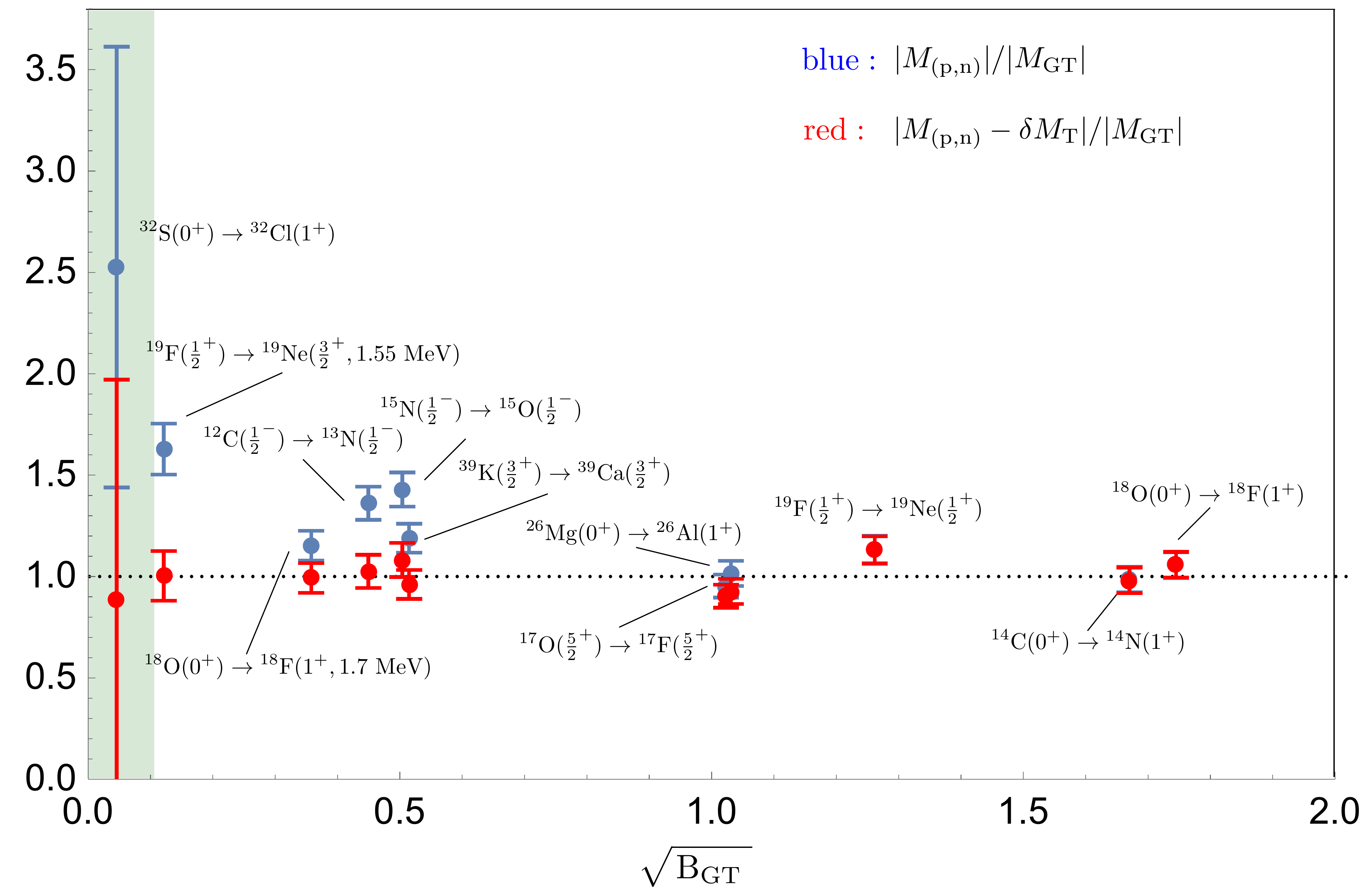}
  \caption{In blue: correspondence between the (p,n) amplitude $|M^\mathrm{(p,n)}|$ and the beta decay amplitude $|M_\mathrm{GT}|$
is excellent when B$_\mathrm{GT}$ is strong, but deteriorates for weaker B$_\mathrm{GT}$.  In red: the agreement is restored with the inclusion of $|M_\mathrm{T}|$.
The two excited states that contribute to the BEST cross section have weak transition strengths that would place them in the shaded region,
where the agreement between (p,n) cross sections and B$_\mathrm{GT}$ is typically poor unless the tensor correction is included.}
  \label{fig:Tensor}
\end{figure}

The results, displayed in Fig. \ref{fig:Tensor}, are quite dramatic.  A naive use of (p,n) reactions to map B$_\mathrm{GT}$ works well for strong transitions,
but deteriorates as the transition strengths lessen, becoming highly unreliable for transitions of the strength of current interest, given by Eq. (\ref{eq:Krofcheck}).
However, with the inclusion of the typically subdominant tensor operator, the correlation between the (p,n) measurements and known weak strengths is restored.  
In using Eq. (\ref{eq:MT}), Ref. \cite{cross_2023} takes into account uncertainties in measurements, in the determination of the strength constant $\delta$, 
and in variations in the theoretical estimates of $M_T$.   The end results is a determination of the tensor strength,
$\delta = 0.075 \pm 0.008$ (1$\sigma$).   

The effective operator can then be used in conjunction with the (p,n) measurements to extract the needed $\mathrm{B_{GT}}$ strengths for \nuc{71}Ga.  The
results can be expressed as \cite{cross_2023}
\begin{equation}
{ \mathrm{B_{GT}}(\textstyle{5 \over 2}^-) \over \mathrm{B_{GT}}(\mathrm{gs})} < 0.089 ~~~(68\% \, \mathrm{C.L.}) ~~~~~~~~~~~~~~
{ \mathrm{B_{GT}}(\textstyle{3 \over 2}^-) \over \mathrm{B_{GT}}(\mathrm{gs})} = 0.121 \pm 0.026~~~(68\% \, \mathrm{C.L.}) 
\label{eq:BGTvals}
\end{equation}
The total cross section can be expressed in terms of these ratios as \cite{Hata1995}
\begin{equation}
 \sigma = \sigma_\mathrm{gs} \left[ 1 + \xi(\textstyle{ 5 \over 2}^-)\, \displaystyle{ \mathrm{B_{GT}}(\textstyle{5 \over 2}^-)  \over \mathrm{B_{GT}}(\mathrm{gs}) }+ \xi(\textstyle{ 3 \over 2}^-)\,\displaystyle{ \mathrm{B_{GT}}(\textstyle{3 \over 2}^-)  \over \mathrm{B_{GT}}(\mathrm{gs}) } \right]
 \end{equation}
 where the phase-space coefficients are \cite{cross_2023}
\[ ^{51}\mathrm{Cr}:~~ \xi(\textstyle{ 5 \over 2}^-)=0.669~~~\xi(\textstyle{ 3 \over 2}^-)=0.220~~~~~~~~~~^{37}\mathrm{Ar}:~~ \xi(\textstyle{ 5 \over 2}^-)=0.696~~~\xi(\textstyle{ 3 \over 2}^-)=0.264  \]
This then leads to the following total cross sections
\begin{eqnarray}
\sigma(^{51}\mathrm{Cr}) &=& \left\{ \begin{array}{ll} 5.71^{+0.27}_{-0.10}  &   68\% \, \mathrm{C.L.} \\[.2cm]    5.71^{+0.51}_{-0.23}  &   95\%~\mathrm{C.L.}  \end{array} \right\}   \times 10^{-45}  
 \,\mathrm{cm}^2\nonumber \\
 & & \\
 \sigma(^{37}\mathrm{Ar}) &=&  \left\{ \begin{array}{ll} 6.88^{+0.34}_{-0.13}  &   68\% \, \mathrm{C.L.} \\[.2cm]    6.88^{+0.63}_{-0.28}  &   95\%~\mathrm{C.L.}  \end{array} \right\}  \times 10^{-45} \,\mathrm{cm}^2 \nonumber
 \label{eq:K}
\end{eqnarray}

This analysis leads to somewhat larger excited-state contributions because the GT and tensor operators interfere destructively for these transitions, 
which then requires a larger $M_\mathrm{GT}$ to compensate.  The excited states increase the total \nuc{71}Ga$(\nu_e,e^-)$\nuc{71}Ge cross sections
by $\sim$ 6.0\% and $\sim$ 6.6\% for \nuc{51}Cr and \nuc{37}Ar, respectively.

Reference \cite{cross_2023} also includes an analysis of excited-state contributions based on forward-angle ($^3$He,t) charge-exchange cross sections \cite{Frekers2015}.

 \subsection{Comparisons to Past Work}
 There are a number of older cross section estimates available in the literature, summarized in Table \ref{tab:CrossSectionEstimates}.  Some of their
 relevant attributes include:
 \begin{enumerate}
 \item  Bahcall (1997) \cite{Bahcall1997}:  This work included, in its estimate of $\sigma_{gs}$, overlap and exchange atomic effects, and used the then
 prevailing value of $Q_\mathrm{EC}$ = 232.69 $\pm$ 0.15 keV.   Excited state B$_\mathrm{GT}$ values were taken from the (p,n) values of \cite{Krofcheck} without
 any added corrections.
\item Haxton (1998) \cite{Haxton1998}:  Extending the arguments of \cite{Hata1995},  this paper was to explore in detail the possible consequences
 of interfering GT and tensor contributions to the (p,n) cross section for the 175 keV excited state, in extracting the neutrino absorption cross section.   The broad error assigned reflects two factors,
 a 30\% larger value for $\delta$, which can be traced to deficiencies in the (p,n) data then available (see \cite{cross_2023}), and a SM estimate of the tensor amplitude $M_T$ for the
 ${3 \over 2}^- \rightarrow {5 \over 2}^-$ transition that was nearly half of the single-particle $2p_{3/2} \leftrightarrow 1f_{5/2}$ value.  As the tensor and GT amplitudes interfere, 
 this allows for a large GT matrix element.  The paper discusses the importance of using the
 full $2p_{3/2}1f_{5/2}1p_{1/2}1g_{9/2}$ SM space in order to properly describe the shape co-existence properties of Ga and Ge, but only the simplest effects of the $1g_{9/2}$ shell were included
 in the calculations done, as
 the technology of the time limited bases to dimensions $\lesssim 10^6$.  (Use of the full SM space generates m-scheme bases exceeding $10^8$).  Consequently, anticipating that the absence of correlation
 would make the truncated SM estimate of $|M_T|$ too large, the author treated the SM estimate as an upper
 bound, yielding the broad range of cross sections shown in Table \ref{tab:CrossSectionEstimates}.
 \item Barinov {\it et al.} (2018) \cite{Barinov2018}:  This work used weak couplings updated to 2018,  including a value for $g_A$ of 1.272 $\pm$  0.002, and adopted the $Q_\mathrm{EC} =233.5 \pm 1.2$ keV,
 which came from a Penning trap measurement of the mass difference \cite{Frekers2013}, though this value had been superseded by a more accurate trapping result from \cite{Alanssari2016}.
 This choice of $Q_{EC}$ accounts in part for the slightly larger cross section obtained, compared to \cite{Bahcall1997}.
 \item Kostensalo {\it et al} (2019) \cite{Kostensalo2019}:  This work followed \cite{Haxton1998} by including the tensor correction in its analysis of excited-state contributions.
 The analysis employs SM estimates of the GT and tensor matrix elements.  In \cite{cross_2023} it is noted that calculations using the identical
 interaction did not reproduce the tabulated GT and tensor matrix elements of \cite{Kostensalo2019}.  
 \item Semenov (2020) \cite{Semenov2020}:  This work follows \cite{Bahcall1997} quite closely, treating the excited states as was done there, but utilizing updated weak couplings and 
 and taking $Q_{EC}$ from \cite{Alanssari2016}, which remains the best value.
 \item Haxton {\it et al.} (2023) \cite{cross_2023}:  As described here, this work includes a much more advanced extraction of the needed excited-state contributions, propagating all identified experimental and theoretical errors
 in the determination of $\delta$ and estimation of $M_T$.   That is, the excited-state treatment follows the  plan of \cite{Haxton1998}, but with improved data and error propagation and without SM limitations.
 Current Particle Data Group and Perkeo II weak couplings were used, and both radiative corrections and the contributions of weak magnetism were included
 in the EC and $(\nu_e,e^-)$ calculations. 
 \end{enumerate}
  
The original BEST analysis \cite{BESTprl,BESTprc} was done using the cross section from Bahcall \cite{Bahcall1997}, which is in reasonably good agreement with the
recent determination of \cite{cross_2023}.  In \cite{cross_2023}, a variety of small changes -- updated values for weak couplings and for $Q_{EC}$, the inclusion of radiative corrections,
the inclusion of weak magnetism, and the computation of Coulomb corrections using a realistic charge distribution consistent with the \nuc {71}Ga r.m.s. charge radius --
combine to lower $\sigma_{gs}$ by about 2.5\%, while the excited state contribution increases to a bit over $\sim$6\% when the effects of $M_T$ are included in the extraction
of B$_\mathrm{GT}$ from (p,n) cross sections.  The net result is a total cross section $\sigma$ $\sim$1.5\% smaller than that of \cite{Bahcall1997}.  For the $\nu_e \rightarrow \nu_s$
oscillation bounds we derive later in this paper, the updated cross section of \cite{cross_2023} is used.
 
\begin {table}[tp]
\caption {A summary of the published neutrino reaction cross section estimates for $^{71}$Ga($\nu_e,e^-$)$^{71}$Ge in units of $10^{-45}$cm$^2$.  The value for $Q_{EC}$ used in each calculation is shown. All results are given at 68\% C.L.  See text for details.  }
\label{tab:CrossSectionEstimates} 
\begin{center} 
 \begin{tabular}{ l c c c c  } 
 \hline\hline
 Author 							& Year	&  $\sigma$(\nuc{51}{Cr})	& $\sigma$(\nuc{37}{Ar}) 		& 	$Q_\mathrm{EC}$(\nuc{71}Ge)  (keV)					\\ [0.5ex]  
 \hline
Bahcall~\cite{Bahcall1997}			&1997	&$5.81^{+0.21}_{-0.16}$ 	& $7.00^{+0.49}_{-0.21}$					&	232.69(15)				\\
Haxton~\cite{Haxton1998}				& 1998	&$6.39\pm0.68$ 					& 		--							&	232.69(15)						\\
Barinov \textit{et al.}~\cite{Barinov2018}	 & 2018 	&$5.91\pm0.11$			& $7.14\pm0.15$			 		 &	233.5(1.2)					\\
Kostensalo \textit{et al.}~\cite{Kostensalo2019} & 2019	& $5.67\pm0.06$			& $6.80\pm0.08$					&	232.49(22)						\\
Semenov~\cite{Semenov2020}		& 2020		& $5.94\pm0.12$			&$7.17\pm0.15$			 		&	232.44(9)				\\
Haxton \textit{et al.}~\cite{cross_2023}    & 2023            & 5.71$^{+0.27}_{-0.10}$       &      6.88$^{+0.34}_{-0.13}$      &      232.44(9)                          \\
\hline\hline
\end{tabular}
\end{center}
\end{table}
	
\subsection{Nuclear and Atomic Data Uncertainties}\label{sec:NPInput}
Various atomic and nuclear parameters are needed in the \nuc{51}Cr and \nuc {37}Ar cross section calculations for \nuc{71}Ga$(\nu_e,e^-)$\nuc{71}Ge, and as noted at various points 
above, small changes have occurred in their values, as new measurements became available over the years.  Here we briefly comment on these data uncertainties, emphasizing that they are quite modest and thus 
cannot account for the observed discrepancy in $R$.\\
~\\	
\noindent
{\it Nuclear Q Values:}~	
The EC Q-values for $^{37}$Ar, $^{51}$Cr, and $^{71}$Ge and their uncertainties are, from the most recent available data evaluation \cite{BNL},
\begin{equation}
 Q_{EC}(^{37}\mathrm{Ar}) = 813.87 \pm 0.20 \mathrm{~keV}~~~~~~Q_{EC}(^{51}\mathrm{Cr}) = 752.39 \pm 0.15 \mathrm{~keV}~~~~~~Q_{EC}(^{76}\mathrm{Ge}) = 232.47 \pm 0.09 \mathrm{~keV} 
 \end{equation}
 The uncertainties range from 0.02\% to 0.04\%, implying an impact on rates or cross sections of $\sim$ 0.04\% to 0.08\%.  This is far below levels of current concern.\\
 ~\\
{\it The \nuc{71}{Ge} half-life:}~The \nuc{71}{Ge} half-life used in the BEST, GALLEX/GNO, and SAGE analyses is $11.43\pm0.03$~d.  This value and its 0.3\% uncertainty comes from~\cite{hampel1985}. The 
neutrino capture cross section depends directly on this value 
as described above.  A recent paper~\cite{Giunti2023} questioned the reliability of this half-life.  Their suggestion of a longer half-life and much larger uncertainty for the \nuc{71}{Ge} EC half-life
was based on four selected measurements, two of which were performed nearly 70 years ago.  One of these older papers drives the conclusions of~\cite{Giunti2023}.
The arguments of \cite{Giunti2023} reflect a misunderstanding of both \cite{hampel1985} and more generally of how nuclear data are evaluated and utilized.  The lifetime of \cite{hampel1985} represents not one measurement,
but the combined results of six distinct measurements that were performed by the authors using different source preparation methods and two distinct counting techniques, in contrast to older work that the authors of \cite{Giunti2023} weighted
equally.  (The authors of \cite{hampel1985} were of course aware of older efforts, noting the age of these measurements as motivation for their efforts to modernize the EC measurement.)
Perhaps more important, the entire body of data on this decay -- which includes \cite{hampel1985} and ten other publications listed in the ENDSF files of the National Nuclear Data Center -- was recently 
re-evaluated \cite{BNNDC}.  The evaluation included
publications as of January 17, 2023, all of which would have have been critically assessed.  The resulting recommended value is also $11.43\pm0.03$~d \cite{BNNDC}. 
Consequently, the speculations of \cite{Giunti2023} are not supported by evaluation of the existing body of data on this decay.  This said, new measurements of this important EC half-life would of course be welcome,
provided they are of the quality of those reported in \cite{hampel1985}. \\
 ~\\
{\it The \nuc{51}{Cr} $\gamma$-decay Branching Ratio:}~
The most precise method of determining \nuc{51}{Cr} activity is calorimetry. Calorimetry requires knowledge of the energy release per decay ($\kappa$), excluding the unmeasured neutrino contribution. The value $\kappa$=37.750$\pm$0.084 keV/decay~\cite{Veretenkin_2017} is dominated by
EC to the first excited state of \nuc{51}V, for which the branching fraction is 9.91$\pm$0.02\%.  The associated $\gamma$ has an energy of 320.0835$\pm$0.0004 keV.   This contribution accounts for 86\% of $\kappa$. As noted by the BEST collaboration~\cite{BESTprc} and studied in Ref.~\cite{Brdar2023}, to first order, if the branching ratio for this decay were incorrect, the deduced source activity would be incorrect by the same factor.  As this branching ratio is known to a precise 0.2\%, this is also not an uncertainty of concern.  Furthermore, any change would not affect the \nuc{37}{Ar} result.\\
~\\
{\it The \nuc{71}{Ge} Electron Density at the Nucleus:}~
The dominant ground-state cross section for \nuc{71}Ga$(\nu_e,e^-)$\nuc{71}Ge is tightly constrained by the known EC rate of \nuc{71}Ge.   This connection
was exploited by Bahcall \cite{Bahcall1997} and by almost all subsequent investigators.   The newest cross section calculation \cite{cross_2023} has now taken into account
radiative corrections and nuclear operator contributions beyond the allowed approximation (dominated by the
interference term with weak magnetism).  These corrections do alter the relationship between EC and $(\nu_e,e^-)$, but enter at the level of $\lesssim 0.5$ \%.  
In addition, this relationship depends on important input from atomic theory,
the $1s$ atomic wave function probability averaged over the nuclear transition density (see discussion in \cite{cross_2023}).  
This averaging generates the probability $|\phi_{1s}^{av}|^2$ that appears in Eq. (\ref{eq:sigmao}).   Uncertainties in $|\phi_{1s}^{av}|^2$ directly
impact the calculated EC rate.

In \cite{Bahcall1997} Bahcall cites as private communications three relativistic, self-consistent Hartree-Fock calculations that took into
account the finite extent of the nucleus, the Breit interaction, vacuum polarization, and self-energy corrections, averaging the resulting wave functions over
the nuclear volume to obtain $|\phi_{i}^{av} |^2$ for $K$, $L$, and $M$ capture.    The calculations, performed by independent groups,
agree at the $\pm 0.2\%$ level.   We are not aware of any subsequent calculations that
are as complete.  While \cite{Bahcall1997} includes references to the atomic methods employed by the three groups, details on the specify calculations performed for $^{71}$Ge
do not appear to be published.  This is somewhat unfortunate, given the importance of $|\phi_{i}^{av} |^2$ in the \nuc{71}Ge EC calculation.

The relationship between the dimensionless numerical quantity given
in \cite{Bahcall1997} and the density $|\phi_{1s}^{av}|^2$ is not entirely obvious: see \cite{cross_2023} for discussion.  When
converted to more conventional units, one finds the result
 \begin{eqnarray}
 ~~~~(\hbar c)^3 |\phi_{1s}^{av} |^2 &=&  (7.21 \pm 0.03) \times 10^{-4} \mathrm{~MeV}^3 \nonumber \\
 &=& \mathcal{R} {(Z \alpha m_e c^2)^3 \over \pi}\Big|_{Z=32}
 \label{eq:dens2}
 \end{eqnarray}
with $\mathcal{R}= 1.333$. The 0.4\% uncertainty (95\% C.L.) is determined from the standard deviations of the three
atomic calculations reported in \cite{Bahcall1997} and from differences in theoretical estimates of overlap and the exchange corrections,
as computed by Bahcall and Vatai (see \cite{cross_2023}).  These two sources of theoretical uncertainty were combined in quadrature.  This 
procedure accounts for differences apparent from the spread among competing calculations,
but not those that could arise if the calculations being compared employed common but flawed
assumptions.   But unless some major mistake has been made in the atomic physics, atomic uncertainties 
are far below the level of current concern.

In the second line of Eq. (\ref{eq:dens2}) the result has been re-expressed in terms
of the Schr\"{o}dinger density for an electron bound to a point charge $Z$, evaluated at the origin.   The dimensionless proportionality factor $\mathcal{R}$
is not too different from unity.\\
~\\	 
{\it K, L, and M Capture Ratios:}~Additional atomic data input uncertainties enter through the experimental K, L, and M EC probabilities
for the \nuc{51}Cr and \nuc{37}Ar sources, listed in Table \ref{tab:SourceIsotopes}.  We see that the absolute branching ratios are known to a typical accuracy of 
$\sim$$10^{-4}$.  Further, any error in these quantities would simply redistribute strength over an atomic energy scale, further diluting any impact.
Consequently these uncertainties are far below levels of concern.

The K, L, and M EC probabilities for \nuc{71}Ge appear in the theoretical expression for the capture, Eq. (\ref{eq:sigmao}), used in \cite{cross_2023}
and in the neutrino analysis presented here.  The uncertainties in these quantities, as well as in the exchange and overlap corrections one needs to
relate theoretical instantaneous EC rates to the physical rates observed in \nuc{71}Ga, are described in \cite{cross_2023}.  This constitutes the dominant
uncertainty in extracting $\mathrm{B_{GT}}(gs)$ from the EC rate.  This uncertainty is propagated into the cross section calculation and reflected in the 
1.5\% uncertainty (95\% C.L.) assigned to $\sigma_{gs}$.  See \cite{cross_2023} for details.
	
\section{The Ga Anomaly and its Possible Implications for Sterile Neutrinos}\label{sec:Oscillations}
The preceding two sections summarize the steps taken to cross-check BEST and earlier Ga calibration experiments.  Despite
a great deal of effort, no candidate explanation has been found involving either a flaw in experimental procedures or uncertainties in the
theoretical input used in the extraction of the \nuc{71}Ga($\nu_e,e^-$)\nuc{71}Ge rate.  Indeed, the experiments are
unusually free of both neutrino source and detector cross section uncertainties.  The sources generate simple line neutrino spectra, calorimetry and other
techniques tightly constrain source intensity, and the known EC rate of \nuc{71}Ge establishes a minimum value for the neutrino
capture cross section on \nuc{71}Ga.

It is possible that the Ga anomaly is a statistical fluctuation -- though a highly improbable one, if all uncertainties have been correctly estimated.  The published BEST results for the inner and outer volumes \cite{BESTprl,BESTprc}, using the 1997 Bahcall 
neutrino absorption cross section, can be compared with those obtained using the updated cross section of \cite{cross_2023},
\begin{equation}
\begin{array}{l}\displaystyle{ { R_\mathrm{out} \over R_\mathrm{out}^\mathrm{expected}}} = 0.77 \pm 0.05      \\
 ~~~ \\[-.4cm]
\displaystyle{ { R_\mathrm{in} \over R_\mathrm{in}^\mathrm{expected}}}  = 0.79 \pm 0.05        \end{array}  ~~~\Rightarrow~~~
\begin{array}{l}\displaystyle{ { R_\mathrm{out} \over R_\mathrm{out}^\mathrm{expected}}} = 0.78 \pm 0.05       \\
 ~~~~ \\[-.4cm]
\displaystyle{ { R_\mathrm{in} \over R_\mathrm{in}^\mathrm{expected}}}  = 0.80 \pm 0.05       \end{array}
\end{equation}
Similarly, for the original Ga anomaly obtained from
the weighted average of the four calibration experiments \cite{abdurashitov2009measurement}
\begin{equation}
{ R \over R^\mathrm{expected}}\Big|_\mathrm{calibration} = 0.87 \pm 0.05     ~~~\Rightarrow~~~ { R \over R^\mathrm{expected}}\Big|_\mathrm{calibration} = 0.88 \pm 0.05
\end{equation}
When all of these data are combined with appropriate ratings weighting, one finds for the updated cross section
\begin{equation}
{ R \over R^\mathrm{expected}}\Big|_\mathrm{combined} = \left\{ \begin{array}{ll} 0.82 \pm 0.03 & \mathrm{uncorrelated} \\ 0.81 \pm 0.05  & \mathrm{correlated} \end{array} \right.
\end{equation}
depending on whether one assumes the uncertainties in the five $^{51}$Cr experiments are uncorrelated or correlated. The dominant correlated uncertainty is that associated with the cross section.
While the original Ga anomaly had a significance of about 2.2$\sigma$, using the updated cross section, with the inclusion of the BEST results that has grown to $\sim 4\
\sigma$ under conservative assumptions.
These estimates are based on our current best knowledge of all input experimental and theoretical uncertainties.
It cannot be attributed to nuclear physics uncertainties in the capture cross section:  Using only capture to the \nuc{71}Ge ground state, one obtains
\begin{equation}
{ R \over R^\mathrm{expected}}\Big|_\mathrm{combined}^\mathrm{minimum~cross~section} = \left\{ \begin{array}{ll} 0.87 \pm 0.03 & \mathrm{uncorrelated} \\ 0.87 \pm 0.05  & \mathrm{correlated} \end{array} \right.
\end{equation}
reducing the significance of the anomaly to approximately $2.6 \sigma$ under the most conservative assumptions, but not eliminating it.

The BEST results have been attributed to \oscil\, but the absence of any distance dependence, from comparing rates in the inner and outer volumes, means that there is no
direct evidence supporting this hypothesis.   The rates observed in the two volumes were each low and consistent within their 1$\sigma$ uncertainties. 

But if \oscil\ is invoked to account for Ga anomaly,  one can check the consistency of this hypothesis with other experiments.  There exist both null results constraining the properties of sterile neutrinos,
and other experimental anomalies that have been linked to their existence.
For a recent review, see Ref.~\cite{Acero2022}.  As discussed in the introduction to this paper, this is not an easy task as the number, masses, and
couplings of possible sterile neutrinos are among the variables one should consider.  Furthermore, sterile neutrinos can be accompanied by other new physics.  As is apparent from Refs. \cite{Acero2022,Adhikari2016,Abazajian2012},
the modeling possibilities have resulted in an extensive literature, much of it generated in the last few years. 

\begin{figure}[t]
    \centering
  \includegraphics[width=0.95\columnwidth]{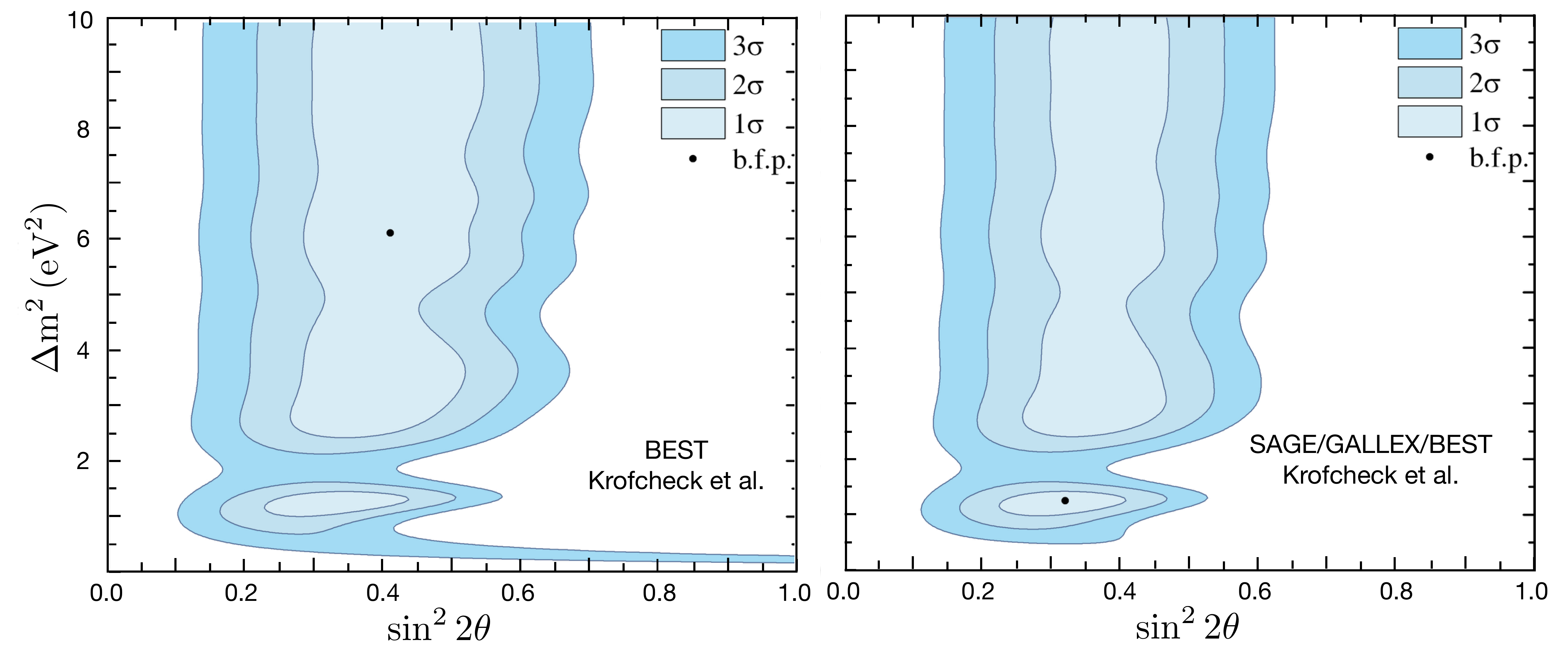}
    \caption{Left: The allowed region for oscillations into a sterile state determined from the BEST inner and outer results, using the update neutrino capture cross section from \cite{cross_2023}. The best-fit point is  sin$^2 2\theta=0.41$,  $\Delta m^2=6.1$~eV$^2$, denoted above by b.f.p. Right: Allowed regions when the constraints from the two GALLEX and two SAGE calibration experiments are added. The best-fit point is sin$^2 2\theta=0.32$, $\Delta m^2 = 1.25$ eV$^2$.  
    The parameter space, however, is very flat. Figure courtesy of Tanya Ibragimova.}
    \label{fig:oscillation}
\end{figure}

Figure~\ref{fig:oscillation} shows the BEST constraints on the simplest (3+1) \oscil\ scenario involving a single sterile state, as well as the constraints when BEST is combined with the SAGE and GALLEX calibration experiments. 
The updated cross section of \cite{cross_2023} has been used.  The parameter space is very flat, particularly along the $\Delta m^2$ direction.  The contours exclude the origin and thus are consistent with the assumption
that \oscil\ is occurring. 

One can then consider whether other results support or are in tension with the hypothesis of \oscil\, for the parameters indicated in Fig. \ref{fig:oscillation}.   BEST's inner/outer detector geometry corresponds to oscillation lengths
corresponding to $\Delta m^2 \sim 1$ eV$^2$, which one sees reflected in Fig.~\ref{fig:oscillation}.  Values much smaller that 1 eV$^2$, corresponding to longer oscillation lengths, are excluded by BEST's reduced counting rate, $R \sim 0.8$.
For $\Delta m^2 \gtrsim 2$~eV$^2$, BEST looses sensitivity to $\Delta m^2$, as the oscillation length is short relative to detector dimensions.  Only the average oscillation is relevant.  Consequently BEST results are compatible
with a wide range of relatively heavy sterile neutrinos.  In contrast, because $R$ is significantly less than 1, a relatively large mixing angle is indicated.  This creates tension with other experimental constraints on sterile neutrinos.

A number of other experiments have produced results impacting the sterile neutrino interpretations of the Ga anomaly: 
	DANSS~\cite{DANSS2021},
	Prospect~\cite{prospect2021},
	St\'{e}r\'{e}o~\cite{stereo2020},
	RENO \& NEOS~\cite{atif2022search,Choi2020RENO} and
	KATRIN~\cite{KATRIN2022}
	all quote limits and provide exclusion regions. As a collective they exclude most, but not all, of the BEST allowed space. The reactor anti-neutrino anomaly~(RAA)~\cite{chooz2011reactor}, and the reactor experiment,
	Neutrino-4~\cite{Neutrino4_2021}, claim evidence for \oscil. The allowed regions for Neutrino-4 and BEST overlap. The allowed regions for RAA and BEST overlap near sin$^2 2\theta \sim$0.2, but marginally. Similarly, limits from
	solar neutrinos~\cite{Giunti2021}
	exclude almost all of the BEST allowed region, except for the lowest allowed mixing angles. The joint
	MiniBooNE--MicroBooNE results~\cite{Aguilar2022mini_microBoone} yield an allowed region that overlaps poorly with the Ga results. Although the MicroBooNE results are limited by low statistics and hence do not significantly alter the MiniBooNE exclusion region, taken by themselves they are consistent with the Ga data~\cite{Denton_2022}.  Readers can find in Ref. \cite{Acero2022} various exclusion plots summarizing existing constraints on the 3+1 sterile neutrino scenario. 
			
On balance no clear evidence has emerged from these experiments that supports the simplest new-physics hypothesis of \oscil\ to a fourth sterile neutrino state as an explanation for the BEST results.  Of course, this does not exclude 
more complicated scenarios with additional beyond-the-Standard-Model degrees of freedom.
On the other hand, as we have described in this review, the many cross-checks of the experimental procedures have been made, yielding no evidence 
of significant issues in either the BEST experiment or the four earlier Ga calibration efforts.  Nor is there any identified theory uncertainty that could possibly account for a $\sim$20\% reduction in the counting rate.  Thus at this time we lack an
explanation for the results that have been obtained.
		
\section{Summary and Outlook}\label{sec:sum}
The BEST experiment was designed as a test of the Ga anomaly that would achieve a higher counting rate, by using a \nuc{51}Cr source of unprecedented intensity to expose a large mass of Ga metal to the neutrino flux.
The two-volume design provided sensitivity to oscillation lengths provided $\Delta m^2 \sim 1$ eV$^2$.   No baseline dependence was observed.  In both volumes a counting rate was obtained that was $\sim$80\% of that expected,
a result consistent with the earlier calibration experiments while also strengthening the statistical signifance the Ga anomaly.

In this review we described the experimental procedures of BEST and the earlier calibration experiments, emphasizing the detailed checks that have been made to verify Ge extraction, proportional counter efficiency, and analysis procedures.  
We discussed the atomic physics of the sources and the multiple checks that have been made of source intensity.  We described the nuclear physics of the \nuc{71}Ga$(\nu_e,e^-)$\nuc{71}Ge cross section, which is tightly constrained by
the known EC rate for \nuc{71}Ge, including recent work that has provided a solid basis for estimating the uncertainty of the remaining $\sim$6\% contribution from \nuc{71}Ge excited states.  Two effects that can alter the
relationship between \nuc{71}Ge EC and neutrino capture on \nuc{71}Ga, radiative corrections and weak magnetism, have been evaluated and found to enter at the level of $\lesssim$0.5\%.  No conventional explanation of the anomaly has been identified, apart from the possibility of
an unfortunate statistical fluctuation.   While it is clearly not possible to rule out some undiscovered experimental artifact
altering either observed rates or current estimates of uncertainties, there is a marked contrast between
the various efficiency tests performed, which typically verified procedures at the level of 1\%, and the $\sim$20\% counting deficit found in the BEST experiment.

The lack of more conventional explanations for the anomaly has led to suggestions that new physics might be at play, specifically an oscillation into a fourth sterile neutrino \oscil.  The BEST and SAGE/GALLEX calibration results are
consistent with such an explanation for a broad range of $\Delta m^2 \gtrsim 1$ eV$^2$ and large mixing angles in the range $\sin^2{2 \theta} \sim 0.3-0.4$.  While sterile neutrinos have been invoked to account for other anomalies,
in general the oscillation parameters indicated by BEST lead to conflicts with various short-baseline null experiments.   While the 3+1 scenario explored is simple and many other possibilities exist, in our view one would need additional supporting
evidence before claiming \oscil\ as a likely solution to the Ga anomaly.

The most productive path forward might be to perform another high-intensity source experiment, improving the precision and helping to further rule out the possibility that the Ga anomaly is totally or partially a statistical fluctuation.
Given the success in producing one high-intensity $^{51}$Cr source (3.14 MCi), one has confidence that a second could be fabricated.

But there are alternatives, including one experiment that would be sensitive to somewhat shorter oscillation lengths, corresponding to higher neutrino mass differences. 
Probing shorter oscillation lengths by making the inner volume of the BEST configuration smaller would be impractical, but developing a higher-energy neutrino source is an intriguing alternative. \nuc{65}{Zn} is an EC isotope with a small $\beta^+$ decay branch (1.421\%). Roughly half the electron captures are to an excited state at 1115.5~keV with the remainder to the ground state with a Q-value of 1352.1~keV. This results in K-capture $\nu_e$s of energies 1342.4~keV and 226.9~keV, the latter below the \nuc{71}{Ga} threshold. Therefore 48.35$\pm$0.11\% of \nuc{65}{Zn} decays produce \nuel\ that can interact.  The longer half-life of \nuc{65}{Zn} (244.01$\pm$0.09~d~\cite{TabRad_v0}) means that many more extractions can be done,
compared to the \nuc{51}{Cr} and \nuc{37}{Ar} source experiments previously performed~\cite{Gorbachev2019}. Furthermore, the \nuc{65}Zn cross section is about three times larger. Even though only 48\% of the decays produce useful $\nu_e$, the count rates would be higher. A first assessment of source fabrication indicates that with 6-7~kg of enriched \nuc{64}{Zn}, a 0.5~MCi source could be produced.   

However, \nuc{65}Zn neutrinos can populate higher energy excited states in \nuc{71}{Ge} at 708~keV, 808~keV and~1096 keV in addition to the states at 175 and 500~keV that contributed to the \nuc{51}{Cr} source experiment.
An estimate of the \nuc{65}{Zn} cross section for \CaptReac\ of (1.82$\pm0.05)\times10^{-44}$~cm$^2$~\cite{Barinov2018} has been made, based on excited-state B$_\mathrm{GT}$ values extracted from forward-angle ($^3$He,t) scattering.
Some 20-30\% of the cross section is due to such states \cite{Barinov2018}.  This is problematic, as the model-dependent contribution to the \nuc{65}Zn \nuc{71}Ga$(\nu_e,e^-)$\nuc{71}Ge cross section would exceed the size of the anomaly one is testing.
In contrast to the (p,n) analysis of \cite{cross_2023}, no systematic effective operator study of ($^3$He,t) as a probe of weak Gamow-Teller strengths has been made.   Thus it is not presently clear whether a \nuc{65}Zn neutrino source experiment
could achieve the precision required, even though certain experimental attributes of this source are attractive.

Huber~\cite{Huber2022GAGG} proposed using the Ce-doped, inorganic scintillating crystal Gd$_3$Al$_2$Ga$_3$O$_{12}$ (Ce:GAGG) to test the anomaly by exposing a 1.5-ton detector of crystals to a BEST-like \nuc{51}{Cr} source ten times.  The charged-current (CC) interaction rate on \nuc{71}{Ga} and the elastic scattering (ES) rate on the electrons within the crystal would both be measured. The ES cross section is well known and therefore the comparison of the two rates is a direct test of the CC cross section. With few previous measurements of CC cross sections in this energy range, this would be a useful measurement even without the motivation of the Ga anomaly.  The absolute activity of the \nuc{51}{Cr} source would cancel out in forming the
ratio of the two rates.  There are clearly advantages to event-by-event detection, compared to the less direct radiochemical method that requires extraction and counting of event products. The CC signature would be an energy deposit of 510 keV ($E_{\nu} - Q$), a number unfortunately near the positron annihilation $\gamma$ energy.  The continuum of ES events extends up to the Q value. A careful background study will be required before the feasibility of this scheme will be known. Ten reproductions of the \nuc{51}{Cr} source would also pose a challenge.

Another possibility would be to place a strong $\bar{\nu}_e$ source near a liquid scintillator detector with position sensitivity and large proton density~\cite{Lasserre2013}. The SOX collaboration~\cite{Gaffiot_2015} had planned to place a $\sim$500~PBq \nuc{144}{Ce}-source near the Borexino detector. The $\bar{\nu}_e$ spectrum from this $\beta$ decay extends up to  3.0~MeV, well above the 1.806~MeV inverse beta decay (IBD) threshold of hydrogen. The sensitivity of Borexino to IBD and its position sensitivity meant that an oscillation curve could be mapped out. Unfortunately the fabrication of the source failed~\cite{Cartlidge2018}, causing the experiment to be abandoned. 

The line neutrinos produced in EC combined with calorimetry and other methods to measure source intensities to high precision help to make the source experiments described above quite attractive.  Furthermore the cross sections
for the reactions they induce are often more constrained than would be the case for higher energy neutrinos.  In the example we have treated here, 94\% the \nuc{51}Cr neutrino cross section for \nuc{71}Ga($\nu_e,e^-$)\nuc{71}Ge can be determined from
the \nuc{71}Ge EC rate, independent of nuclear models.  Thus the further development of this field is important, given our incomplete knowledge of neutrino physics and the need for high precision tests of neutrino properties.
	
\section*{Acknowledgements}
We thank Hamish Robertson for helpful discussions.  
This work was supported by the Department of Energy, Office of Nuclear Physics under Federal Prime Agreement LANLEM78 (SRE); 
by the Higher Education of Russian Federation under agreement no. 14.619.21.0009 (unique project identifier
no. RFMEFI61917X0009) (VG); 
and by the US Department of Energy under grants DE-SC0004658, DE-SC0015376, and DE-AC02-05CH11231,
the National Science Foundation under cooperative agreement 2020275, and the Heising-Simons Foundation under award 00F1C7 (WH).

%\section*{Author's contributions \textit{(optional section)}} Detailing here the contributions of the authors of the review.

%\bibliography{GaAnomaly}
	
%	\newpage
%	\appendix
%	\renewcommand*{\thesection}{\Alph{section}}
%	
%	\section{Appendices, if necessary}\label{appendix}

\end{document}

%% file: BESTAbbreviations.tex
\def\nuc#1#2{${}^{#1}$#2}

\def\nuel{$\nu_{e}$}			%electron neutrino symbol
\def\nust{$\nu_{s}$}			%sterile neutrino symbol
\def\oscil{$\nu_{e}\rightarrow\nu_{s}$}			%electron neutrino to sterile neutrino transition
\def\CaptReac{$^{71}$Ga$($\nuel,$e^-)^{71}$Ge}
\def\be{\begin{equation}}
\def\ee{\end{equation}}

\newcommand{\cpowten}[2]{$#1\times10^{#2}$}